\documentclass[reprint,superscriptaddress,floatfix,longbibliography,tightenlines,nobibnotes,nofootinbib,amsmath,amssymb,aps,pra]{revtex4-1}
\usepackage{graphicx}
\usepackage{color}
\usepackage{soul}
\usepackage{bm}
\usepackage{dcolumn}
\usepackage{textcomp}
\usepackage{hyperref}
\hypersetup{colorlinks=true,urlcolor= blue,citecolor=blue,linkcolor= blue,bookmarks=true,bookmarksopen=false}
\usepackage{verbatim}
\usepackage{blindtext}
\usepackage{natbib}
\setcitestyle{numbers,square}
\usepackage{makecell}
\usepackage{amsmath}
\DeclareMathOperator{\sgn}{sgn}
\usepackage{amsfonts}
\usepackage{amssymb}
\usepackage{comment}
\usepackage{mathtools}
\usepackage{braket}
\usepackage{soul}
\begin{document}

\title{Tunable topological phases in monolayer Pt$_2$HgSe$_3$ with exchange fields }
\author{Vassilios Vargiamidis}
\email{V.Vargiamidis@warwick.ac.uk}
\affiliation{School of Engineering, University of Warwick, Coventry, CV4 7AL, United Kingdom}
\author{P. Vasilopoulos }
\email{p.vasilopoulos@concordia.ca}
\affiliation{Department of Physics, Concordia University, 7141 Sherbrooke Ouest, Montreal, Quebec H4B 1R6, Canada}
\author{Neophytos Neophytou}
\email{N.Neophytou@warwick.ac.uk}
\affiliation{School of Engineering, University of Warwick, Coventry, CV4 7AL, United Kingdom}

\begin{abstract}

We investigate topological phases of monolayer jacutingaite (Pt$_2$HgSe$_3$) that arise when considering the competing effects of spin-orbit coupling (SOC), magnetic exchange interactions, and staggered sublattice potential $V$. The interplay between the staggered potential and exchange field offers the possibility of attaining different topological phases. By analyzing the Berry curvatures and computing the Chern numbers and Hall conductivities, we demonstrate that the system is time-reversal-symmetry-broken quantum spin Hall insulator when $m_b < \lambda_{so}$, where $m_b$ is the exchange field operating on the bottom Hg sublattice and $\lambda_{so}$ is the intrinsic SOC. For $m_b > \lambda_{so}$ and in the presence of Rashba SOC, we find that the band gap at valley $K$($K^{\prime}$) is topologically trivial (nontrivial) with Chern number $\mathcal{C} = 1$ and valley Chern number $\mathcal{C}_v = -1$, indicating that the system is valley-polarized quantum anomalous Hall insulator. We show that the topology of each valley is swapped (the Chern number becomes $\mathcal{C} = -1$) by reversing the sign of the exchange field. The system transitions to a valley-polarized metal and quantum valley Hall phase as $V$ increases. Along the phase boundaries, we observe a single Dirac-cone semimetal states. These findings shed more light on the possibility of realizing and controlling topological phases in spintronics and valleytronics devices.
	
\end{abstract}

\maketitle
\date{\today}

\section{Introduction}

In the past few years, unprecedented efforts have been devoted to the exploration of novel topological insulator phases and of their remarkable properties at their boundaries \cite{hasan10,qi11}. In two dimensions (2D), their hallmark is the presence of propagating edge states, which carry dissipationless currents. These transport properties depend crucially on the nature of the bulk energy gaps and the associated topological invariants together with the underlying symmetries of the system \cite{bernevig13}. For instance, the quantum anomalous Hall (QAH) effect could arise when time-reversal ($\mathcal{T}$) symmetry is broken by local magnetization \cite{weng15}. The QAH effect is characterized by a nonzero topological invariant $\mathcal{C}$ - known as Chern number - and is similar to the integer quantum Hall (IQH) effect but without Landau-level quantization.

Another topological phenomenon is the quantum spin Hall (QSH) effect, a state of matter originating from spin-orbit coupling (SOC) and the preservation of $\mathcal{T}$ symmetry \cite{kane05,kane05b,bernevig06}. Its nontrivial topology is characterized by a $\mathbb{Z}_2$ index \cite{kane05b} or a spin Chern number \cite{sheng06}; the two descriptions being equivalent for $\mathcal{T}$ invariant systems \cite{prodan09}. The spin Chern number is well defined even if $\mathcal{T}$ symmetry is broken and the QSH state was shown to survive in this case \cite{yang11}. Their topological and dissipation-free transport properties make the QAH and QSH insulators outstanding material platforms for the realization of quantum-based technologies, including spintronics \cite{avsar20} and topological valleytronics \cite{vitale18}.

A major challenge for practical applications is the identification of experimentally synthesized QSH insulators that persist up to room temperature. The QSH phase was first realized in semiconductor quantum wells based on HgTe/CdTe \cite{konig07,konig13} and InAs/GaSb \cite{knez11,suzuki13} heterostructures, and in 2D materials like WTe$_2$ \cite{fei17,wu18,shi19}, while the QAH phase was demonstrated in magnetically (Cr or V) doped (Bi, Sb)$_2$Te$_3$ thin films \cite{chang13,check14,chang15}. However, edge state transport in these systems occurs at low temperatures due to their small bulk energy gaps. The search for topological phases in materials with large gap is thus highly desirable.

Recently, using first-principles simulations, the first large-gap QSH insulator was predicted to be monolayer jacutingaite (Pt$_2$HgSe$_3$) \cite{maraz18}, a new species of platinum-group minerals \cite{cabral08}, which was also synthesized \cite{vyma12}. It has a sandwich-like structure with a platinum (Pt) layer between two selenium (Se) and mercury (Hg) layers. Its Hg atoms at the top and bottom layers form a buckled honeycomb lattice, similar to that in silicene and germanene. Its low-energy physics around the Fermi level can be described by the Kane-Mele (KM) model, originally introduced for graphene \cite{kane05,kane05b}, but with significantly stronger SOC; it gaps the Dirac point making the system an insulator with a band gap of $\approx 0.15$ eV at the DFT level \cite{maraz18,maraz19,bafek20}, and $\approx 0.5$ eV as obtained from many-body $G_0 W_0$ calculations \cite{maraz19}. QSH to QAH phase transition in monolayer jacutingaite was demonstrated by chemical functionalization \cite{luo21}. It was also identified as a promising candidate to realize topological valleytronics when interfaced with a 2D magnet \cite{liu20}, and it can potentially host unconventional superconductivity \cite{fink19}. A strong valley polarization and a layer-type band inversion around a single valley was also found in Pt$_2$HgSe$_3$/CrGeTe$_3$ van der Waals (vdW) heterostructure \cite{rehman22}. Furthermore, bilayer jacutingaite was shown to undergo a topological transition from trivial to QSH insulator upon the application of a perpendicular electric field \cite{rade21}. Some experimental evidence, including high stability in air and measurement of its band gap was reported recently \cite{kand20}. 

Apart from jacutingaite, a large class of highly stable, similar materials hosting the QSH phase based on the KM model was recently investigated \cite{lima20}. Jacutingaite-like materials, such as Pt$_2$HgSe$_3$ and Pd$_2$HgSe$_3$, with broken inversion ($\mathcal{P}$) symmetry were also shown to exhibit several promising valley-spin-based phenomena, such as the coupled spin-valley Hall effect, valley spin-valve effect, and selective excitation of carriers from opposite valleys \cite{rehman22b}.

In this work, we explore topological phase transitions in monolayer Pt$_2$HgSe$_3$ with magnetic exchange field and staggered sublattice potential $V$. The induced inequality of sublattice potential necessarily arises in the presence of a perpendicular external electric field. The proximity-induced exchange field is attainable by placing Pt$_2$HgSe$_3$ on a magnetic substrate; its presence breaks both $\mathcal{P}$ and $\mathcal{T}$ symmetries. It also leads to different exchange fields, $m_b$ and $m_t$, at the bottom and top Hg sublattices due to their unequal separations from the substrate [see Fig.~1(d)]. We consider two regimes: i) $m_b < \lambda_{so}$, and ii) $m_b > \lambda_{so}$, where $\lambda_{so}$ is the intrinsic SOC. In the first regime, we show that for $V = 0$ the system is a $\mathcal{T}$-symmetry-broken QSH insulator. The QSH phase exists up to a certain value of $V$ at which a gap closing and reopening occur and the system enters a valley polarized metal (VPM) phase. Further increasing $V$ drives the system into a quantum valley Hall (QVH) phase. At the phase boundaries, we observe a single Dirac-cone (SDC) semimetal states. We characterize and distinguish different topological phases by computing topological invariants (Chern numbers), and/or by computing the spin- and valley-resolved Hall conductivities. We derive analytical expressions for these conductivities and show that their quantized values persist up to relatively high temperatures of $\sim 90$ K.

In the second regime opposite spin valence and conduction bands in one valley are pushed upward and downward, respectively, and penetrate each other, resulting in spin-mixing. In this case, turning on the Rashba SOC introduces spin-flip terms into the Hamiltonian, and consequently the $z$-component of the spin is no longer conserved. We show that the system is valley-polarized quantum anomalous Hall (VP-QAH) insulator \cite{pan14,jena17} for $V = 0$ with charge Chern number $\mathcal{C} = 1$ and valley Chern number $\mathcal{C}_v = -1$. As $V$ increases from zero, the system remains in the VP-QAH phase for a certain range of $V$, and thereafter it exhibits the VPM and QVH phases. Importantly, we find that reversing the sign of the exchange interaction swaps the topology of each valley and the Chern number becomes $\mathcal{C} = - 1$.

In Sec.~II we present the Hamiltonian of monolayer Pt$_2$HgSe$_3$ and discuss the topological invariants. In Sec.~III, we discuss the topological phases in the regime $m_b < \lambda_{so}$ and characterize them with the Chern numbers and with the spin- and valley-Hall conductivities. In Sec.~IV we discuss the VP-QAH phase. We summarize and conclude in Sec.~V.

\section{Model and topological theory}

The band structure of monolayer Pt$_2$HgSe$_3$ can be described by a tight-binding Hamiltonian constructed from first-principles \cite{maraz18} that shares several terms with the KM model for a QSH insulator in graphene \cite{kane05b}. In the long-wavelength limit at Dirac points $K$ and $K^{\prime}$ to linear order in the relative wave vector $k = ( k_x^2 + k_y^2 )^{1/2}$ it reads
\begin{eqnarray}
\nonumber \hspace*{-0.07in} H_\tau = \hbar v_F \left( \tau \sigma_x k_x + \sigma_y k_y \right) \mathbf{1}_s + \lambda_{so} \tau \sigma_z s_z + m_+ \mathbf{1}_\sigma s_z
\\* &&\hspace*{-2.75in} + m_- \sigma_z s_z + \lambda_R \mathbf{1}_s \left( \tau \sigma_x s_y - \sigma_y s_x \right) + V \sigma_z \mathbf{1}_s  ,
\label{eq1}%
\end{eqnarray}
where $\boldsymbol \sigma$ and $\mathbf{s}$ are Pauli matrices that correspond to the sublattice pseudospin and spin degrees of freedom, respectively, and $\mathbf{1}_\sigma ( \mathbf{1}_s )$ denotes the identity matrix in the $\boldsymbol \sigma ( \mathbf{s} )$ space; $\tau = \pm 1$ labels the valley ($K, K^{\prime}$) degree of freedom. The first term in Eq.~(\ref{eq1}) is a massless graphene-type Hamiltonian with $v_F = 3\times 10^5$ m/s. The second term is the KM SOC with $\lambda_{so} = 81.2$ meV \cite{maraz18}, which is four orders of magnitude larger than that in graphene, and respects $\mathcal{T}$ and $\mathcal{P}$ symmetries. The strong SOC combined with the exchange fields significantly lifts the valley degeneracy and leads to energy gaps of opposite sign at the two valleys, leading to different topological responses. The third and fourth terms represent proximity-induced exchange interactions \cite{zol16} described by $m_{\pm} = \left( m_b \pm m_t \right) / 2$, where $m_+$ and $m_-$ stand for symmetric and anti-symmetric parts of the inequivalent exchange fields $m_b$ and $m_t$. The vertical distance between the two Hg sublattices is $3.49$ \AA ~\cite{bafekry20} and the lattice constant of monolayer jacutingaite is $a = 7.6$ \AA ~\cite{maraz18}. The bottom sublattice experiences much stronger magnetic proximity effect due to the stronger atomic wave function overlap, which makes $m_b$ stronger than $m_t$. The staggered exchange field $m_{-} \sigma_z s_z$ preserves $\mathcal{P}\mathcal{T}$ symmetry (see Appendix A for a formal proof). On the other hand, the sublattice independent Zeeman field $m_{+} s_z$ breaks $\mathcal{P}\mathcal{T}$ symmetry thus lifting the spin degeneracy of the energy bands, as will be illustrated below. The $m_{+}$ field also shifts the Dirac cones so as to break electron-hole symmetry. The fifth term in Eq.~(\ref{eq1}) is the Rashba SOC, $\lambda_R$, and arises due to a perpendicular electric field or interaction with the substrate. The last term arises due to a staggered sublattice potential $V$; it breaks the $\mathcal{P}$ symmetry and lifts the spin degeneracy on its own.

The 2D spinfull model in Eq.~(\ref{eq1}), without the exchange fields and the sublattice potential, was proposed by Kane and Mele \cite{kane05} and constitutes a generic description of the QSH phase. This $4\times4$ matrix model is defined on a honeycomb lattice and is closely related to the tight-binding model applicable to all Xenes, i.e., silicene, germanene, and stanene. The model incorporates the effect of the SOC, of both intrinsic and Rashba types, and is $\mathcal{T}$ symmetric. Even though jacutingaite is a ternary material and is somewhat different from the Xenes, it possesses a buckled honeycomb structure of mercury atoms, which is ultimately responsible for its KM physics. \cite{comment}

\subsection{Eigenvalues and eigenfunctions}

Equation (\ref{eq1}) can be written explicitly as a $4 \times 4$ matrix,
\begin{equation}
H_\tau = \left(
\begin{array}
[c]{cc}%
H_{\tau}^{\uparrow} 
 &  R_\tau \\ \
R_{\tau}^{\dagger}  &  H_{\tau}^{\downarrow}
\end{array}
\right), \label{eq2}%
\end{equation}
where the diagonal elements are given by
\begin{equation}
H_{\tau}^{s_z} = \left(
\begin{array}
[c]{cc}%
m_b s_z + \Delta_{\tau}^{s_z} 
 &  \hbar v_F ( \tau k_x - i k_y ) \\
\hbar v_F ( \tau k_x + i k_y )  &  m_t s_z - \Delta_{\tau}^{s_z}
\end{array}
\right), \label{eq3}%
\end{equation}
with $\Delta_{\tau}^{s_z} = V + \lambda_{so} \tau s_z$ and $s_z = \uparrow, \downarrow$ for the  spin-up ($\uparrow$) and spin-down ($\downarrow$) states that correspond to $s_z = \pm 1$. We assume that $m_t < m_b$. This is a reasonable assumption because the exchange fields induced by the hybridization between Pt$_2$HgSe$_3$ and substrate orbitals become weaker with increasing separation. We also assume $0 < m_{-} < \lambda_{so}$. The off-diagonal elements are due to the Rashba interaction,
\begin{equation}
R_{\tau} = \left(
\begin{array}
[c]{cc}%
0 
 &  -i \lambda_R ( \tau - 1 ) \\
i \lambda_R ( \tau + 1 )  &  0
\end{array}
\right) . \label{eq4}%
\end{equation}
We first diagonalize the Hamiltonian in the absence of Rashba interaction, $\lambda_R = 0$, which is relevant for $m_b < \lambda_{so}$, and then for $\lambda_R \neq 0$, which is for $m_b > \lambda_{so}$.\\
\begin{figure}[t]
\vspace*{0.2cm}
\begin{center}
\includegraphics[height=6.25cm, width=8.6cm ]{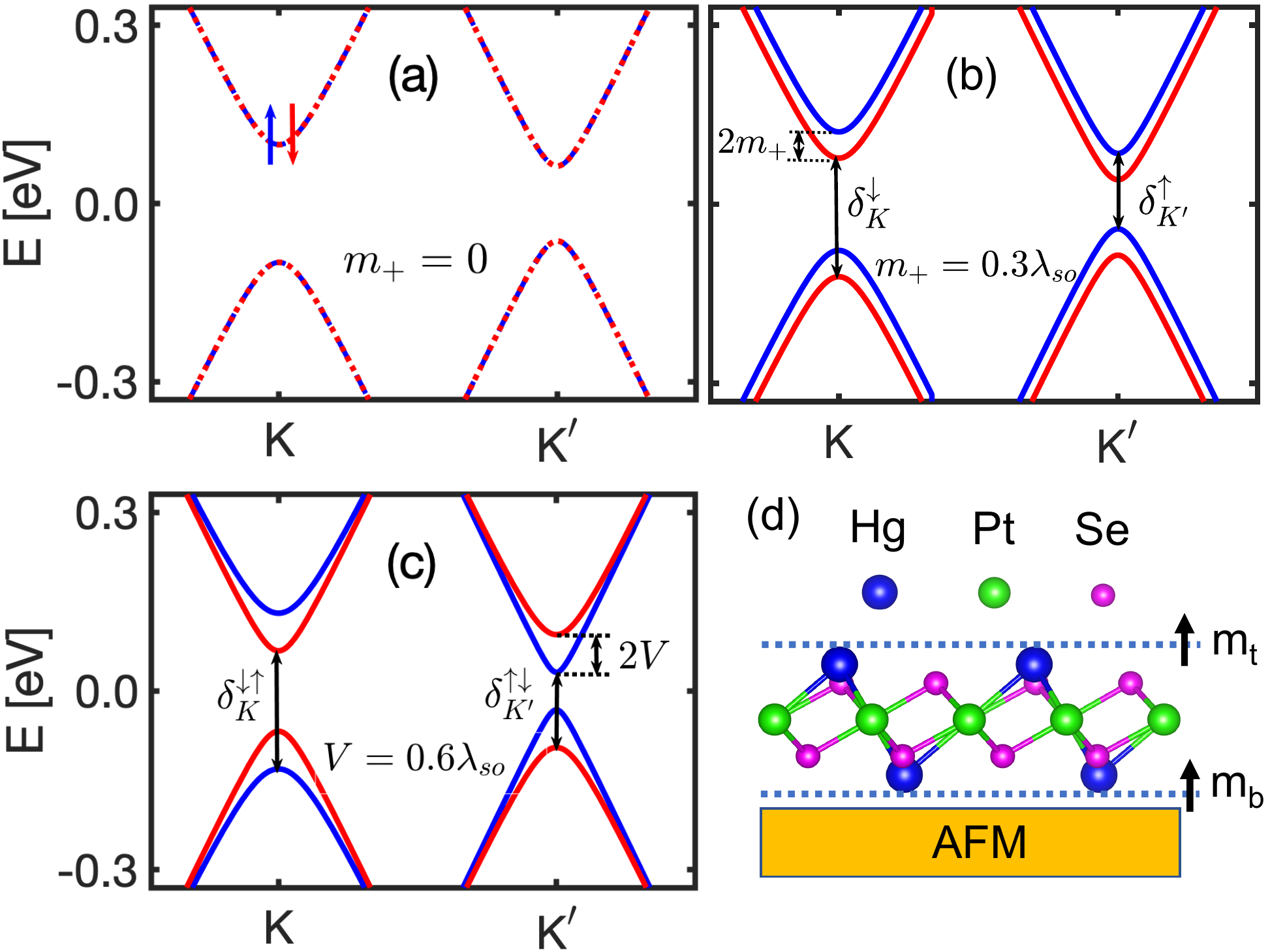}
\end{center}
\vspace*{-0.3cm} \caption{(Colour online) (a) The band structure of monolayer Pt$_2$HgSe$_3$ near the $K$ and $K^{\prime}$ valleys in the absence of the Zeeman field $m_{+}$ and $V=0$. (b) The same as in (a) but with $m_{+}=0.3 \lambda_{so}$. The blue (red) lines are for spin up (spin down) states. The splitting of the spin-degenerate bands is $2 m_{+}$. (c) The same as in (a) but with $V = 0.6 \lambda_{so}$. The splitting of the spin-degenerate bands is $2 V$. (d) Side view of the structure of Pt$_2$HgSe$_3$ on an antiferromagnetic (AFM) substrate. The exchange fields $m_b$ and $m_t$ are shown by the arrows.  }
\label{fig:fig1}%
\end{figure} 

{\bf i) Case  ${\mathbf{\lambda_R = 0}}$}.

The spin-up and spin-down matrices in the Hamiltonian (\ref{eq2}) can be diagonalized separately. The diagonalization of Eq.~(\ref{eq3}) gives the energy dispersion,
\begin{equation}
E_{n s_z k} = m_+ s_z + n 
\big[M_{\tau s_z}^2 + \epsilon_k^2 \big]^{1/2} ,
\label{eq5}%
\end{equation}
where $\epsilon_k = \hbar v_F k$, $M_{\tau s_z} = m_- s_z + \Delta_{\tau}^{s_z}$, and $n = \pm$ stands for the conduction ($+$) and valence ($-$) bands. In Fig.~1(a) we show the energy bands of monolayer Pt$_2$HgSe$_3$ for $m_b = 0.55 \lambda_{so}$, $m_t = 0.05 \lambda_{so}$, and $V = 0$, but we neglect the Zeeman field and set $m_{+} = 0$. For nonzero $m_{+}$ and $V$ the spin splitting between the two-fold degenerate levels is given by $2 (m_{+} + V)$ at $k = 0$, as shown in Figs.~1(b) and 1(c). The gap between spin-up bands at the $K^{\prime}$ valley, $\delta_{K^{\prime}}^{\uparrow} = 2 \vert \lambda_{so} - m_{-} - V \vert$, closes at $V = V_1 = \lambda_{so} - m_{-}$ and reopens for $V > V_1$, as will be illustrated in Sec.~II. The gap between spin-down bands at the $K$ valley, $\delta_{K}^{\downarrow} = 2 \vert \lambda_{so} + m_{-} - V \vert$, closes at $V = V_2 = \lambda_{so} + m_{-}$ and reopens for $V > V_2$. All other gaps remain open. The eigenstates of Eq.~(\ref{eq3}) are $\Psi_{n s_z k} ( \mathbf{r} ) = u_{n s_z k} e^{i \mathbf{k} \cdot \mathbf{r}} / \sqrt{S}$, where $S$ is the area of the sample and $u_{n s_z k}$ are given by
\begin{eqnarray}
\hspace*{-0.6cm} 
u_{n s_z k} = \frac{1}{ 
[ \epsilon_k^2 + (n \varepsilon - M_{\tau s_z} )^2 
]^{0.5}}\,  \Big(
\begin{array}
[c]{cc}%
\epsilon_k \\
\tau 
( n \varepsilon - M_{\tau s_z} 
) e^{i \tau \varphi_k}
\end{array}
 \Big), 
\label{eq6}%
\end{eqnarray}
where $\varepsilon = \left( M_{\tau s_z}^2 + \epsilon_k^2 \right)^{1/2}$ and $\tan \tau \varphi_k = \tau k_y / k_x$.\\

{\bf ii) Case  ${\mathbf{\lambda_R \neq 0}}$}. 

The diagonalization of Eq.~(\ref{eq2}) with $\lambda_R \neq 0$ leads to a quartic equation for the eigenvalues,
\begin{equation}
E^4 +a_1 E^2 + a_2 E + a_3 = 0 ,
\label{eq7}%
\end{equation}
with the coefficients $a_i$ given by
\begin{equation}
a_1 = - m_b^2 - m_t^2 - 2 \lambda_{so} ( \lambda_{so} + 2 m_- \tau ) - 2 V^2 - 2 \epsilon_k^2 - 4 \lambda_R^2 ,
\label{eq8}%
\end{equation}
\vspace{-0.8cm}
\begin{equation}
\hspace{-2.88cm} a_2 = 8 \left( \lambda_R^2 \tau - m_+ V \right) \left( m_- + \lambda_{so} \tau \right)  ,
\label{eq9}%
\end{equation}
\vspace{-0.8cm}
\begin{eqnarray}
\nonumber \hspace*{-1.15cm} a_3 = ( \gamma_1^{\uparrow} \gamma_2^{\uparrow} - \epsilon_k^2 ) ( \gamma_1^{\downarrow} \gamma_2^{\downarrow} - \epsilon_k^2 ) - \lambda_R^2 ( \tau + 1 )^2 \gamma_1^{\uparrow} \gamma_2^{\downarrow}  
\\* &&\hspace*{-6.12cm} - \lambda_R^2 ( \tau - 1 )^2 \gamma_2^{\uparrow} \gamma_1^{\downarrow}  ,
\label{eq10}%
\end{eqnarray}
where $\gamma_1^{s_z} = m_b s_z + \Delta_{\tau}^{s_z}$ and $\gamma_2^{s_z} = m_t s_z - \Delta_{\tau}^{s_z}$. The solutions of Eq.~(\ref{eq7}) are
\begin{eqnarray}
\nonumber \hspace*{-0.2cm} E_{n s k}  &=& \frac{1}{2 \sqrt{3}}  \Big\{ n \Big[ - 2 a_1 + B + \frac{A}{2^{1/3}} \Big]^{1/2} 
- s \Big[  4 a_1 + B\\* 
&& +\frac{A}{2^{1/3}} 
+ n \frac{6 \sqrt{3} a_2}{\sqrt{- 2 a_1 + B + A/2^{1/3}}} \Big]^{1/2} \Big\}  ,
\label{eq11}%
\end{eqnarray}
\begin{figure}[t]
\vspace*{0.2cm}
\begin{center}
\includegraphics[height=6.5cm, width=8.6cm ]{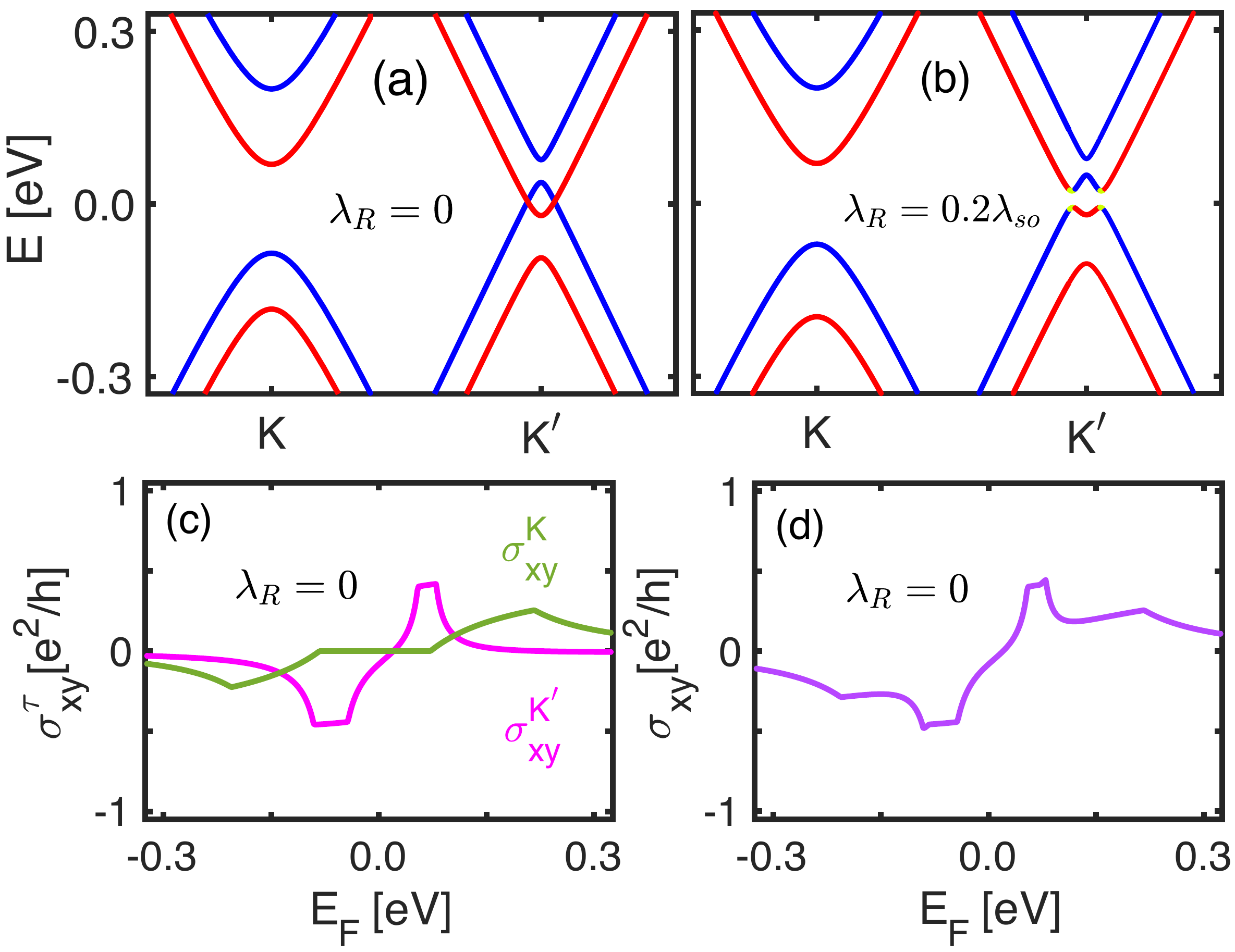}
\end{center}
\vspace*{-0.3cm} \caption{(Colour online) (a) The band structure of monolayer Pt$_2$HgSe$_3$ near the $K$ and $K^{\prime}$ valleys for $m_b = 1.35 \lambda_{so}$, $m_t = 0.05 \lambda_{so}$, $V = 0.3 V_1$, and $\lambda_R = 0$. The band gap $\delta_{K^{\prime}}^{\downarrow\uparrow} < 0$. (b) The same as in (a) but with $\lambda_R = 0.2 \lambda_{so}$. (c) Valley-Hall conductivities $\sigma_{xy}^K$, $\sigma_{xy}^{K^{\prime}}$, see Sec.~IIB. (d) Anomalous Hall conductivity $\sigma_{xy}$ for $\lambda_R = 0$.  }
\label{fig:fig1}%
\end{figure} 
\noindent where $n = \pm$ stands for the conduction ($+$) and valence ($-$) bands, and $s = \pm$ is the spin chirality. $A,\,B,\, C$ are constants. $C=2 a_1^3 + 27 a_2^2 - 72 a_1 a_3  $\,\, and
\begin{eqnarray}
A &=& \Big[ C+
[C^2 - 4 ( a_1^2 + 12 a_3 )^3]^{1/2} \Big]^{1/3} ,\\*
B &=& 2^{1/3} ( a_1^2 + 12 a_3)/A.
\label{eq12}%
\end{eqnarray}
The corresponding eigenstates for the $K$ ($\tau = +1$) and $K^{\prime}$ ($\tau = -1$) valleys are $\Psi_{n s k}^{K, K^{\prime}} ( \mathbf{r} ) = u_{n s k}^{K, K^{\prime}} e^{i \mathbf{k} \cdot \mathbf{r}} / \sqrt{S}$, where $u_{n s k}^{K} $ and $u_{n s k}^{K^{\prime}}$ are given by
\begin{equation}
\hspace{-0.5cm}u_{n s k}^{K} = N \left[ \frac{\epsilon_k e^{-i \varphi_k}}{E_{n s k} - \gamma_1^{\uparrow}} \eta, \eta, \frac{E_{n s k} - \gamma_2^{\downarrow}}{\epsilon_k e^{i \varphi_k}}, 1 \right]^T , 
\label{eq14}%
\end{equation}
\vspace{-0.5cm}
\begin{equation}
\hspace{-0.25cm}u_{n s k}^{K^{\prime}} = \tilde{N} \left[ \tilde{\eta}, -\frac{\epsilon_k e^{-i \varphi_k}}{E_{n s k} - \gamma_2^{\uparrow}} \tilde{\eta}, -\frac{\epsilon_k e^{ i \varphi_k}}{E_{n s k} - \gamma_1^{\downarrow}}, 1 \right]^T  \hspace{-0.2cm } ,
\label{eq15}%
\end{equation}
with $T$ for the transpose; $\eta$ and $\tilde{\eta}$ are given by 
\begin{eqnarray}
\eta &=& \frac{\epsilon_k^2 - ( E_{n s k} - \gamma_1^{\downarrow} ) ( E_{n s k} - \gamma_2^{\downarrow} )}{2 i \lambda_R \epsilon_k e^{i \varphi_{k}}}\, ,\\*
\label{eq16}%
\notag
\ \\
\vspace{0.9cm}
\tilde{\eta} &=& \eta 
\epsilon_k e^{i \varphi_k}/ [E_{n s k} - \gamma_1^{\downarrow}],
\label{eq17}%
\end{eqnarray}
with $N$ and $\tilde{N}$ the normalization constants defined by
\begin{eqnarray}
\nonumber \hspace*{-0.0cm}
N = \Big \{\vert \eta \vert^2 \Big[ 1 + \frac{\epsilon_k^2}{ ( E_{n s k} - \gamma_1^{\uparrow} )^2} \Big] +1 + \frac{ ( E_{n s k} - \gamma_2^{\downarrow} )^2}{\epsilon_k^2}    \Big\}^{-1/2}  .  \\*
\label{eq18}%
\end{eqnarray}
\begin{eqnarray}
\nonumber \hspace*{-0.0cm}
\tilde{N} = \Big \{\vert \tilde{\eta} \vert^2 \Big[ 1 + \frac{\epsilon_k^2}{ ( E_{n s k} - \gamma_2^{\uparrow} )^2} \Big] +1 + \frac{\epsilon_k^2}{ ( E_{n s k} - \gamma_1^{\downarrow} )^2}    \Big\}^{-1/2}  .  \\*
\label{eq19}%
\end{eqnarray}
We note that, for $m_b > \lambda_{so}$, the gap at the $K$ valley remains positive, $\delta_{K}^{\downarrow\uparrow} > 0$, but the gap at the $K^{\prime}$ valley becomes negative, $\delta_{K^{\prime}}^{\downarrow\uparrow} < 0$, for the range $0 < V < m_{+}$ [see Fig.~2(a) and Appandix B]. This leads to spin-degeneracy circles in momentum space at energy $E = 0$. Turning on the Rashba interaction mixes the spin-up and spin-down states, produces an avoided band crossing, and opens a band gap as shown in Fig.~2(b) where we plot the energy dispersion from Eq.~(\ref{eq11}) for $\lambda_R = 0.2 \lambda_{so}$. We argue in Sec.~IV that the gap at the $K$ valley is topologically trivial and that the gap at the $K^{\prime}$ valley is topologically nontrivial and displays the VP-QAH effect. We also show that this topological phase can be switched between valleys by reversing the sign of the exchange interaction.

The presence of the Zeeman field $m_+$ and the sublattice potential $V$ break the $\mathcal{T}$ and $\mathcal{P}$ symmetries leading to nonzero Berry curvature and hence nonzero anomalous Hall conductivity. This is shown in Figs.~2(c) and 2(d). In Fig.~2(c) we show the Hall conductivities for each valley $\sigma_{xy}^{K}$, $\sigma_{xy}^{K^{\prime}}$, see Sec.~IIB, and in Fig.~2(d) the total anomalous Hall conductivity $\sigma_{xy}$ for the same parameters as in Fig.~2(a).

\subsection{Topological invariants}

A bulk gap in the energy spectrum can be characterized by a topological invariant, which is insensitive to deformations of the band structure provided that the gap remains open. Topological invariants to index a topological insulator phase are the charge Chern number and spin Chern number \cite{hasan10,qi11,eza12,niu16}. In the insulating regime when the Fermi level lies in the bulk gap, the Hall conductivity for each spin component is given as
\begin{equation}
\sigma_{xy}^{s_z} = \frac{e^2}{h} \sum_{\tau = \pm} \mathcal{C}_{\tau}^{s_z}  , 
\label{eq20}%
\end{equation}
where $\mathcal{C}_{\tau}^{s_z} $ is the spin- and valley-dependent Chern number for a band which is evaluated as \cite{xiao10,thou82}
\begin{equation}
\mathcal{C}_{\tau}^{s_z} = \frac{1}{2 \pi} \int d^2 k \hspace{0.02cm} \Omega_n^{\tau s_z} ( \mathbf{k} )  .
\label{eq21}%
\end{equation}
In Eq.~(\ref{eq21}) $\Omega_n^{\tau s_z} ( \mathbf{k} )$ is the Berry curvature in the out-of-plane direction for the $n$th band,
\begin{eqnarray}
\hspace*{-0cm} 
\nonumber
\Omega_{n}^{\tau s_z} ( \mathbf{k} ) = - 2 \hbar^2 \text{Im}\sum_{n^{\prime} \neq n } f_{n k} \frac{\langle u_{n k} \vert v_x  \vert u_{n^{\prime} k} \rangle \langle u_{n^{\prime} k } \vert v_y \vert u_{n k} \rangle}{\left( E_{n k} - E_{n^{\prime} k } \right)^2} , \\*
\label{eq22}%
\end{eqnarray}
where $f_{n k}$ is the Fermi function for band $n$, $\vert u_{n k} \rangle$ is the Bloch state with energy eigenvalues $E_{n k}$, and $v_\nu$ ($\nu = x, y$) is the velocity operator; the spin index has been suppressed for brevity. The velocity matrix elements in Eq.~(\ref{eq22}) are evaluated in Appendix C. The total Chern number is then evaluated as $\mathcal{C} = \mathcal{C}_{\uparrow} + \mathcal{C}_{\downarrow}$, where $\mathcal{C}_{\uparrow} = \sum_{\tau} \mathcal{C}_\tau^{\uparrow}$ and $\mathcal{C}_{\downarrow} = \sum_{\tau} \mathcal{C}_\tau^{\downarrow}$ are the Chern numbers for each spin sector. The charge Hall conductivity is given as $\sigma_{xy} = \sigma_{xy}^{\uparrow} + \sigma_{xy}^{\downarrow}$, where the Hall conductivity for each spin component is given in Eq.~(\ref{eq20}). On the basis of $\mathcal{C}_{\tau}^{s_z} $, the spin Chern number is defined as $\mathcal{C}_s = \mathcal{C}_{\uparrow} - \mathcal{C}_{\downarrow}$ \cite{sheng06,niu16,comment2}. The spin-Hall conductivity can then be expressed as $\sigma_{xy}^s = \sigma_{xy}^{\uparrow} - \sigma_{xy}^{\downarrow}$; it is equal to the spin Chern number up to a normalization constant $e^2 / h$. In the presence of spin non-conserving terms, this definition remains valid, but the calculation of $\mathcal{C}_s$ relies on a decomposition of the occupied band into two sectors via diagonalization of the spin operator \cite{prodan09,yang11}. In addition, the valley Chern number is defined as $\mathcal{C}_v = \mathcal{C}_K - \mathcal{C}_{K^{\prime}} = \sum_{s_z} \left( \mathcal{C}_{K}^{s_z} - \mathcal{C}_{K^{\prime}}^{s_z} \right)$ and the Hall conductivity for each valley as $\sigma_{xy}^{\tau} = (e^2 / h) \mathcal{C}_{\tau}$.
\begin{figure*}[t]
\vspace*{0.2cm}
\hspace*{-0.1cm}\includegraphics[height=3cm, width=3.9cm ]{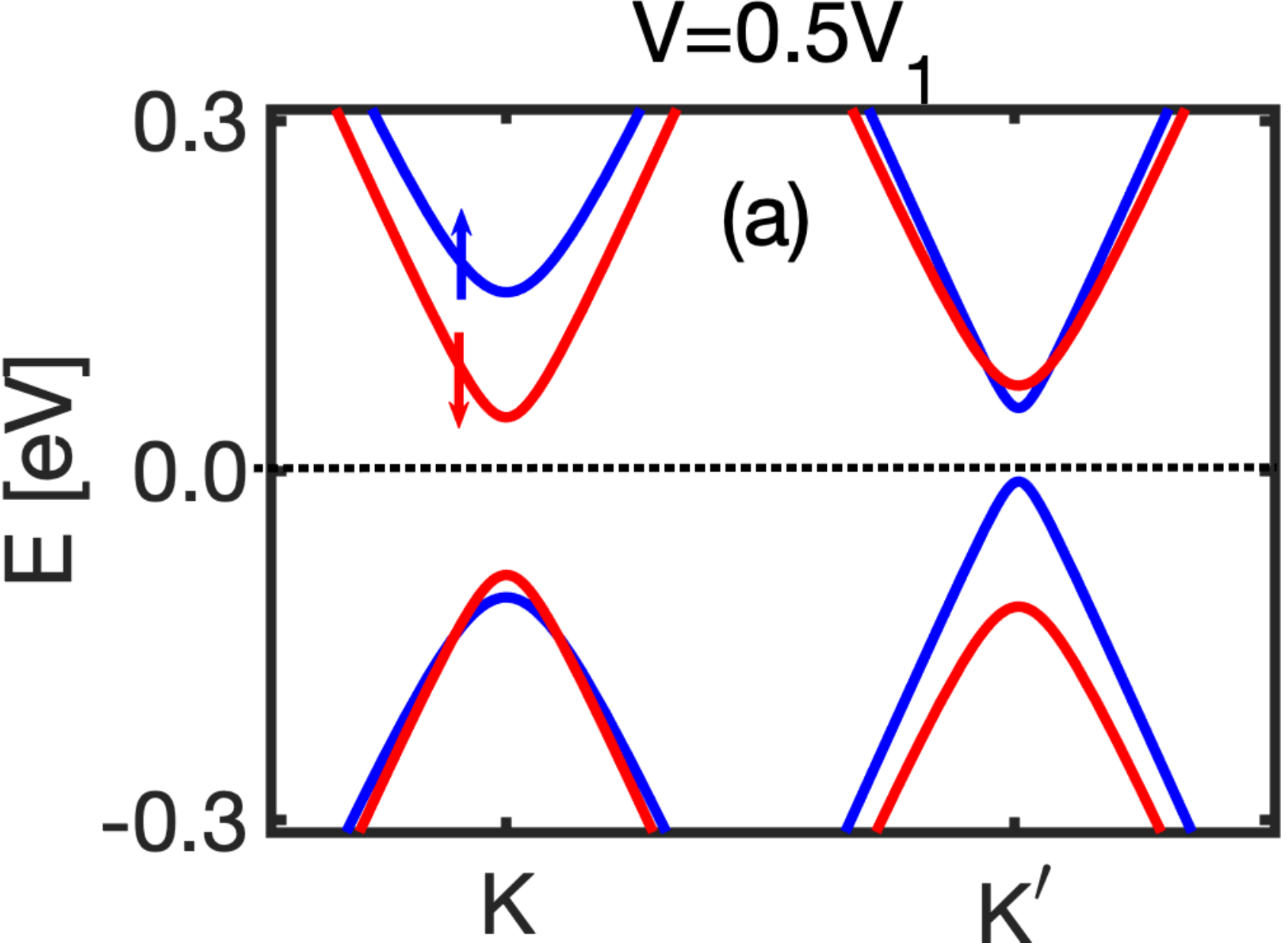}
\hspace{-0.11cm}\includegraphics[height=3cm, width=3.5cm ]{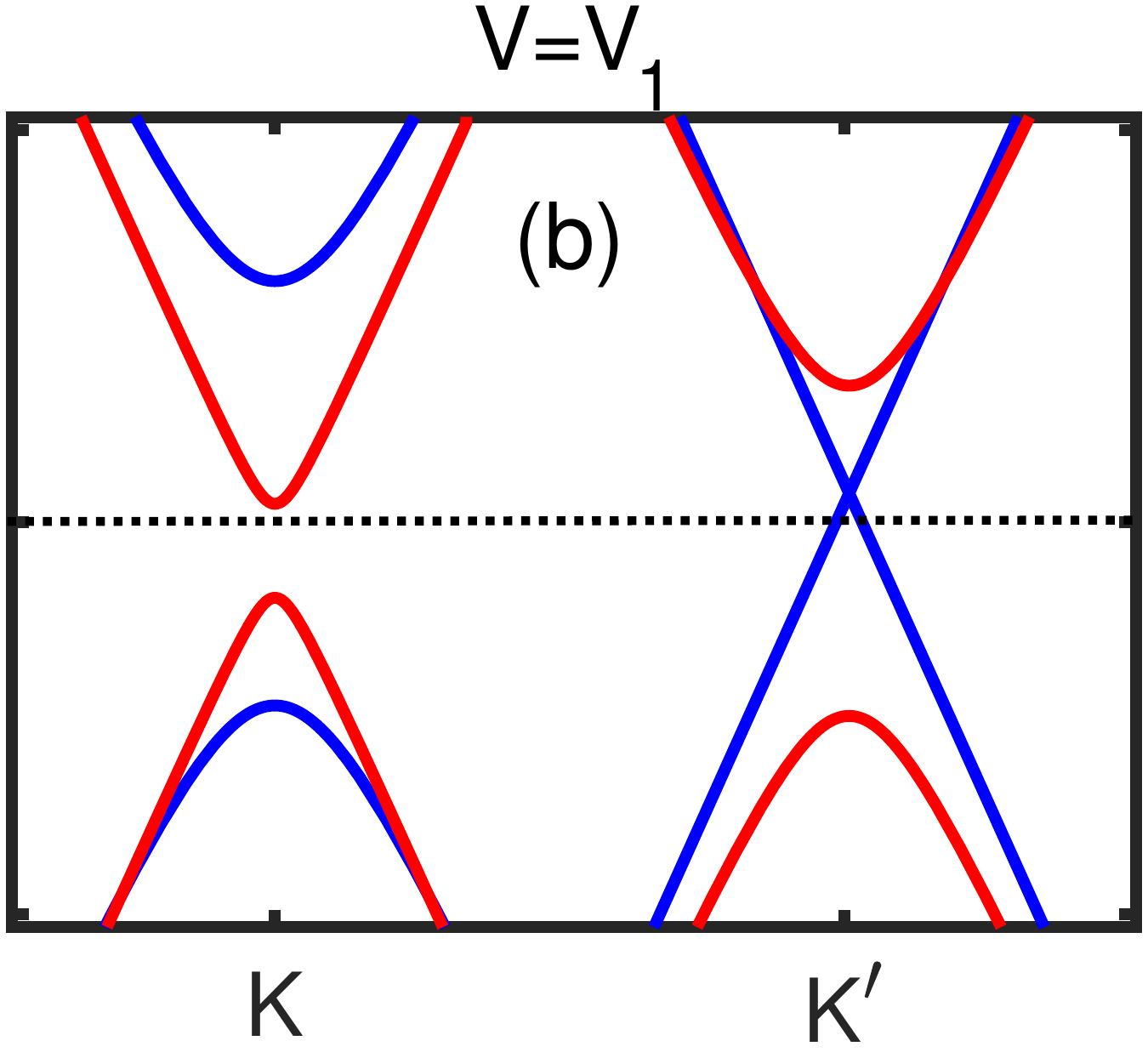}
\hspace{-0.12cm}\includegraphics[height=3cm, width=3.5cm ]{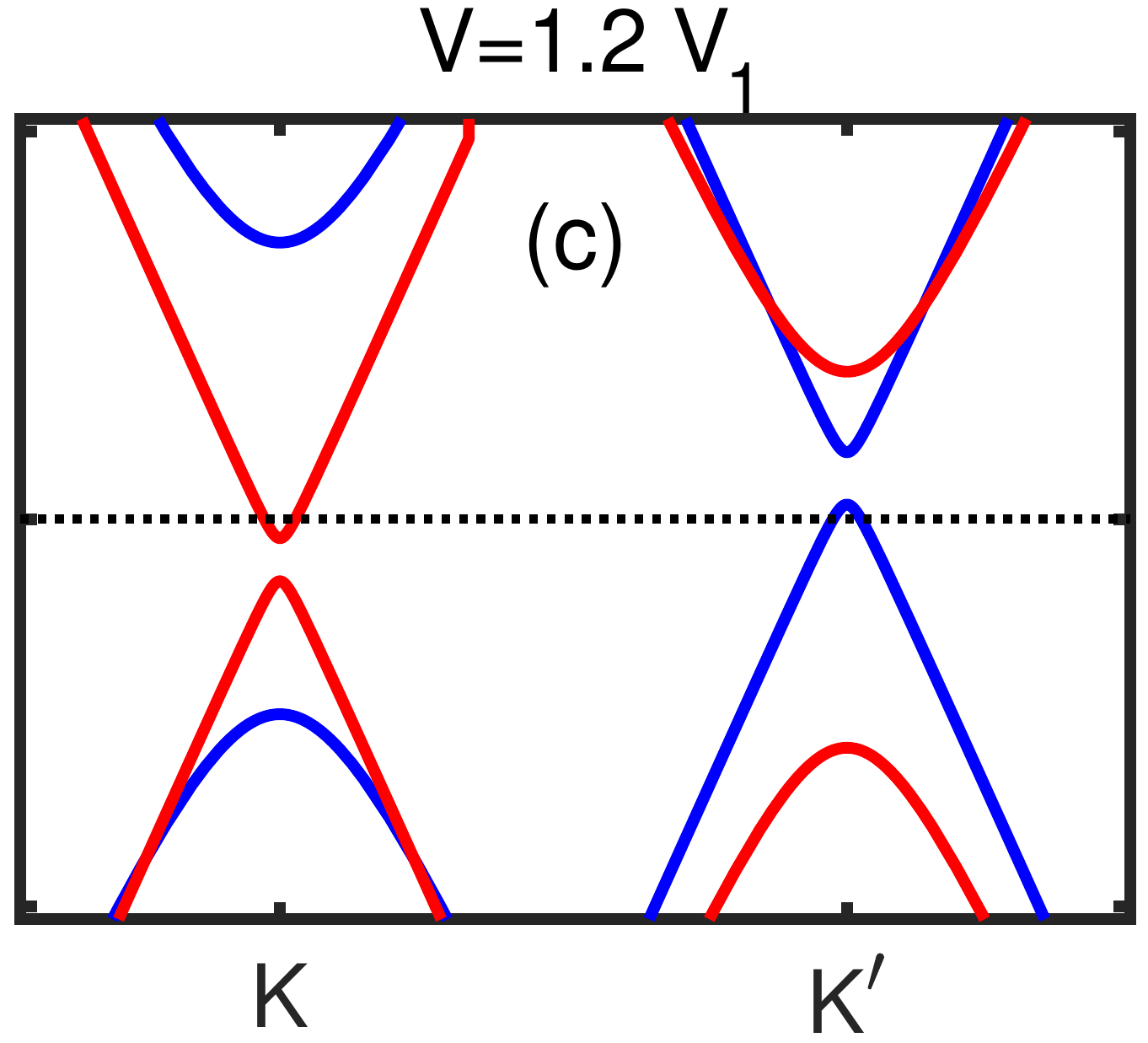}
\hspace{-0.12cm}\includegraphics[height=3cm, width=3.5cm ]{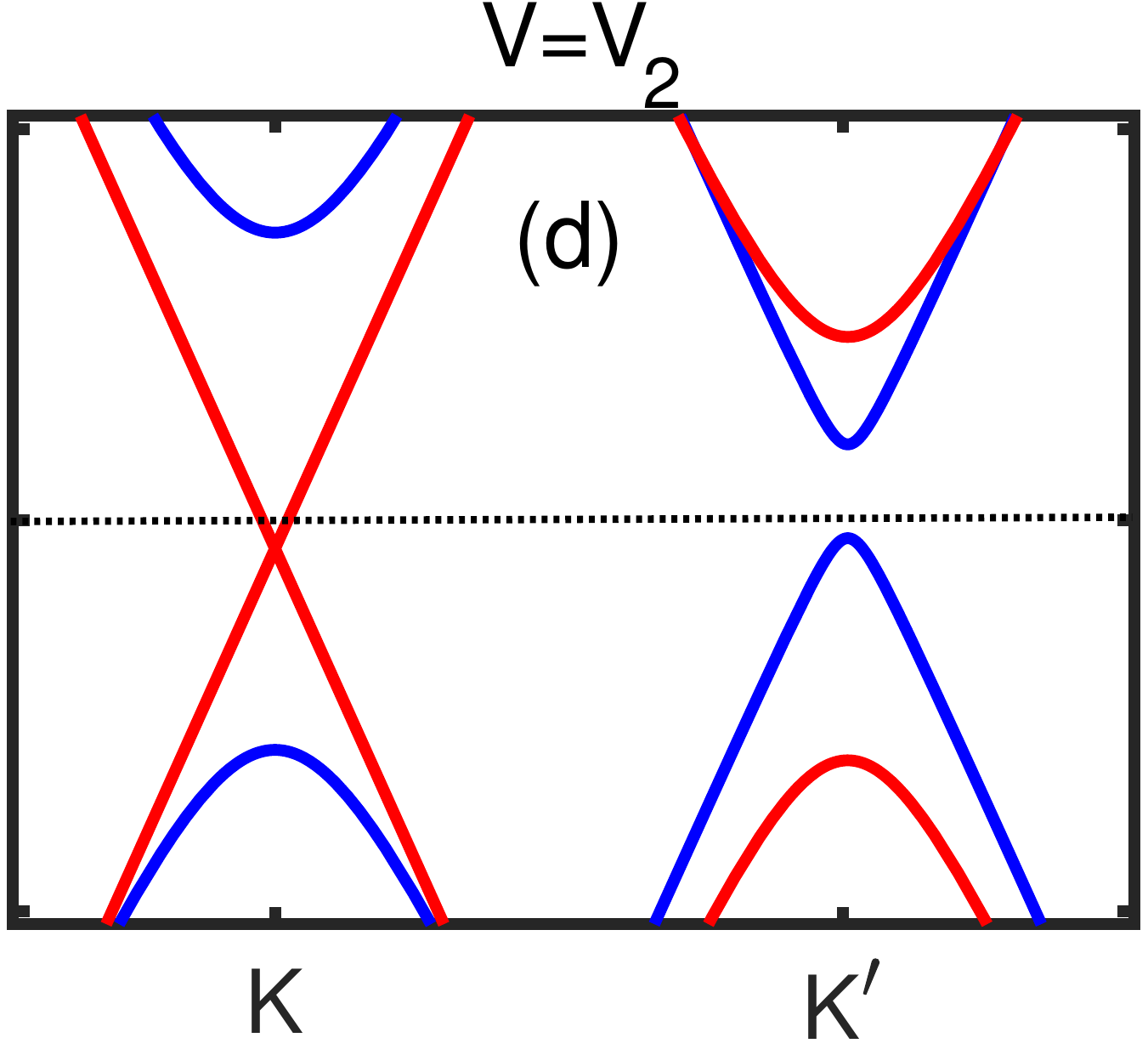}
\hspace{-0.12cm}\includegraphics[height=3cm, width=3.5cm ]{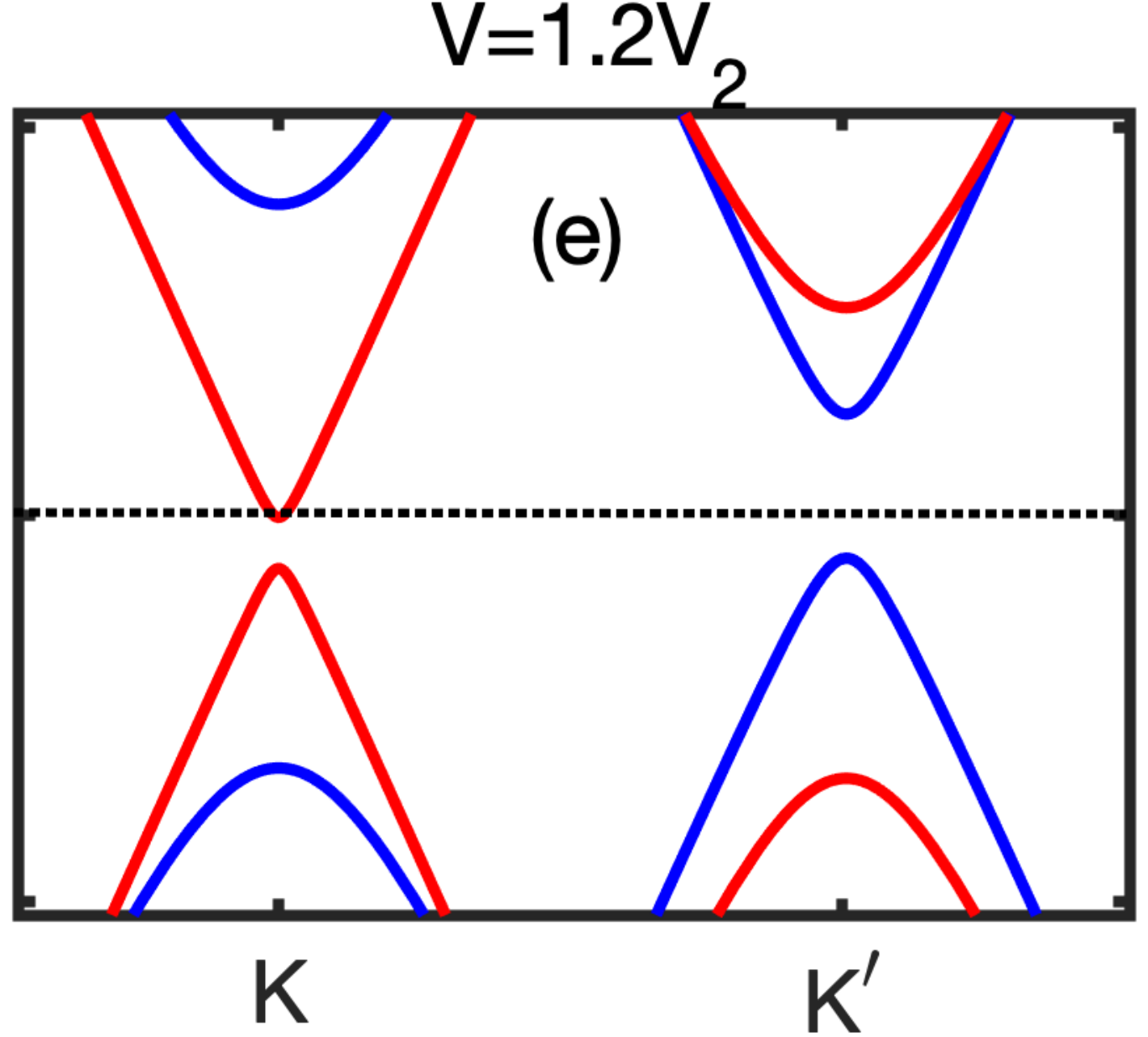}\\
\vspace{0.15cm}
\hspace{-0.45cm}\includegraphics[height=2.9cm, width=4.5cm ]{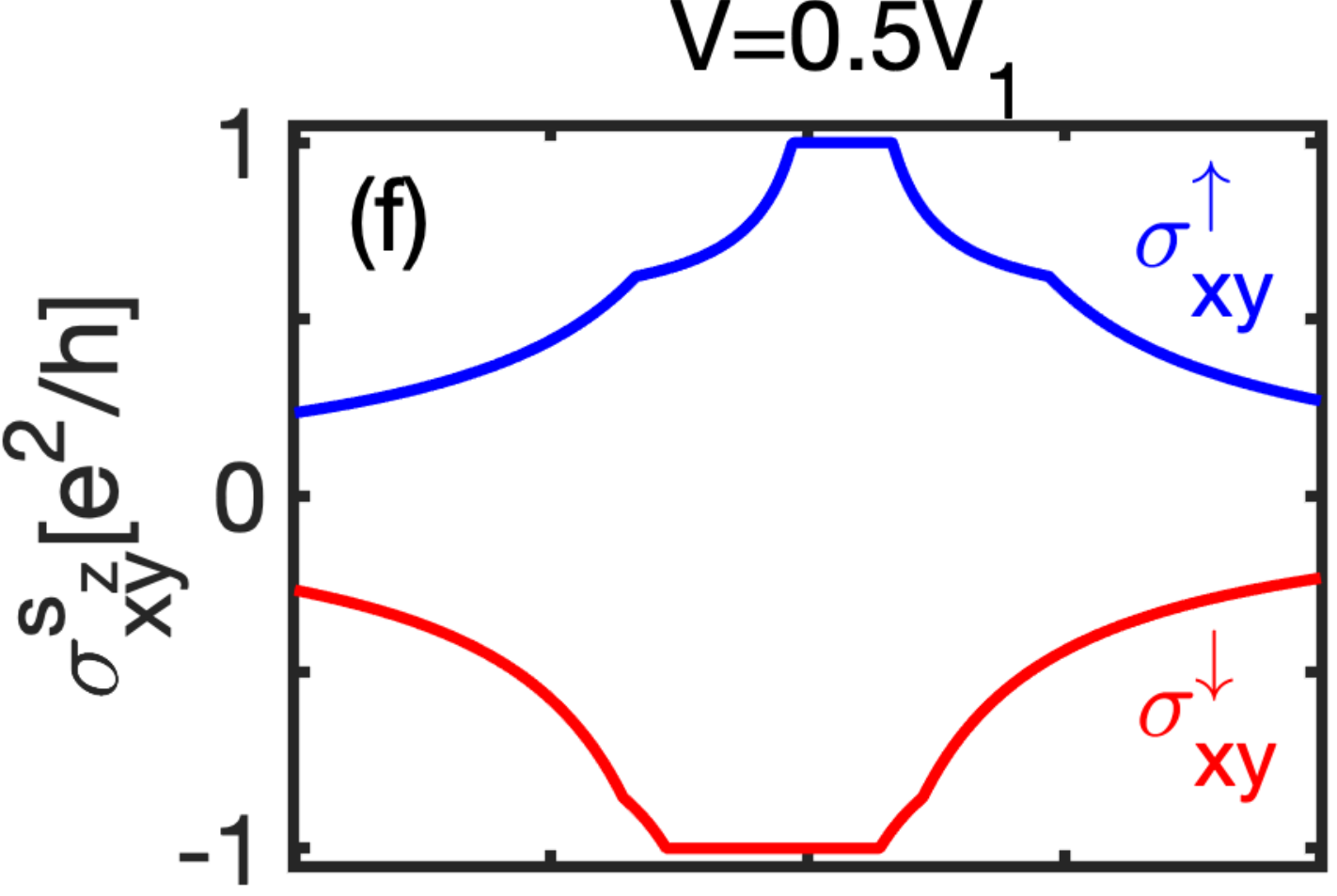}
\hspace{0.02cm}\includegraphics[height=2.9cm, width=3.46cm ]{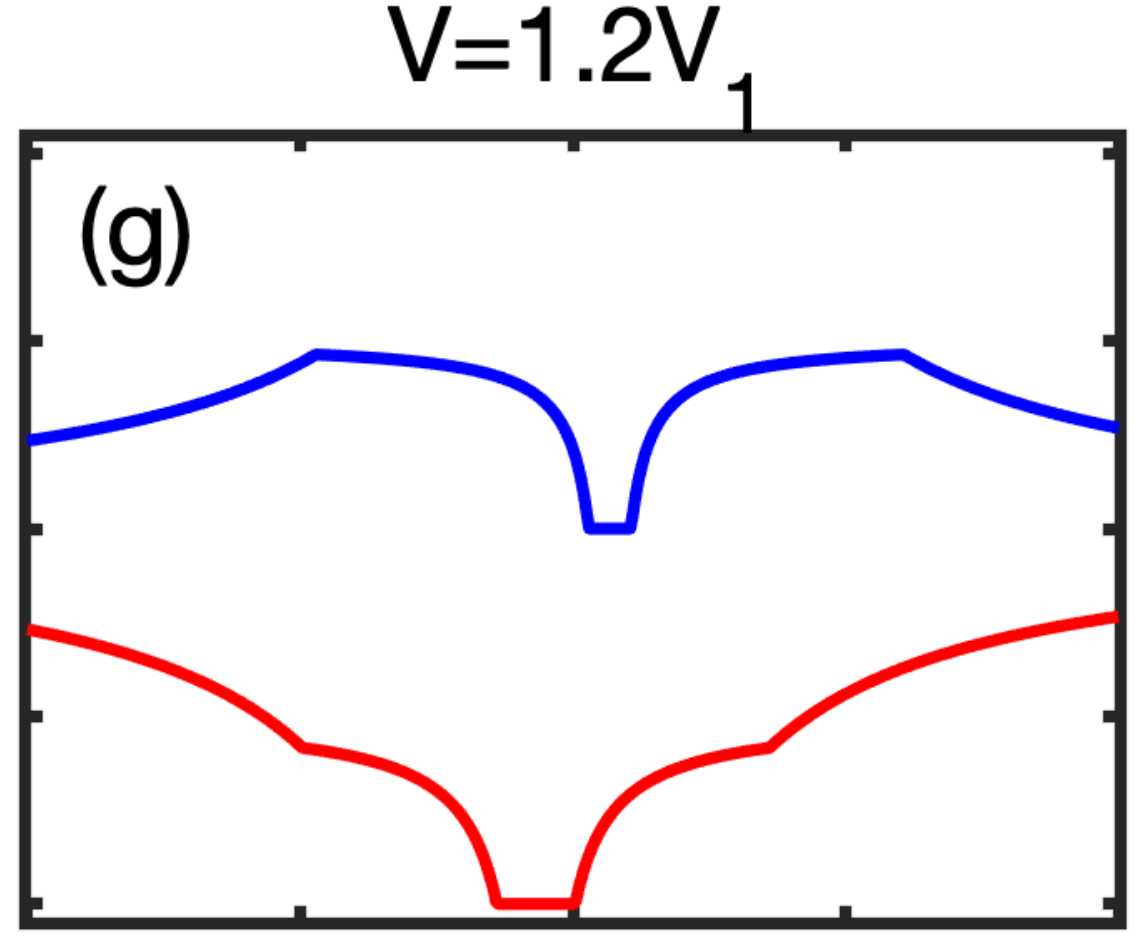}
\hspace{-0.0cm}\includegraphics[height=2.9cm, width=3.46cm ]{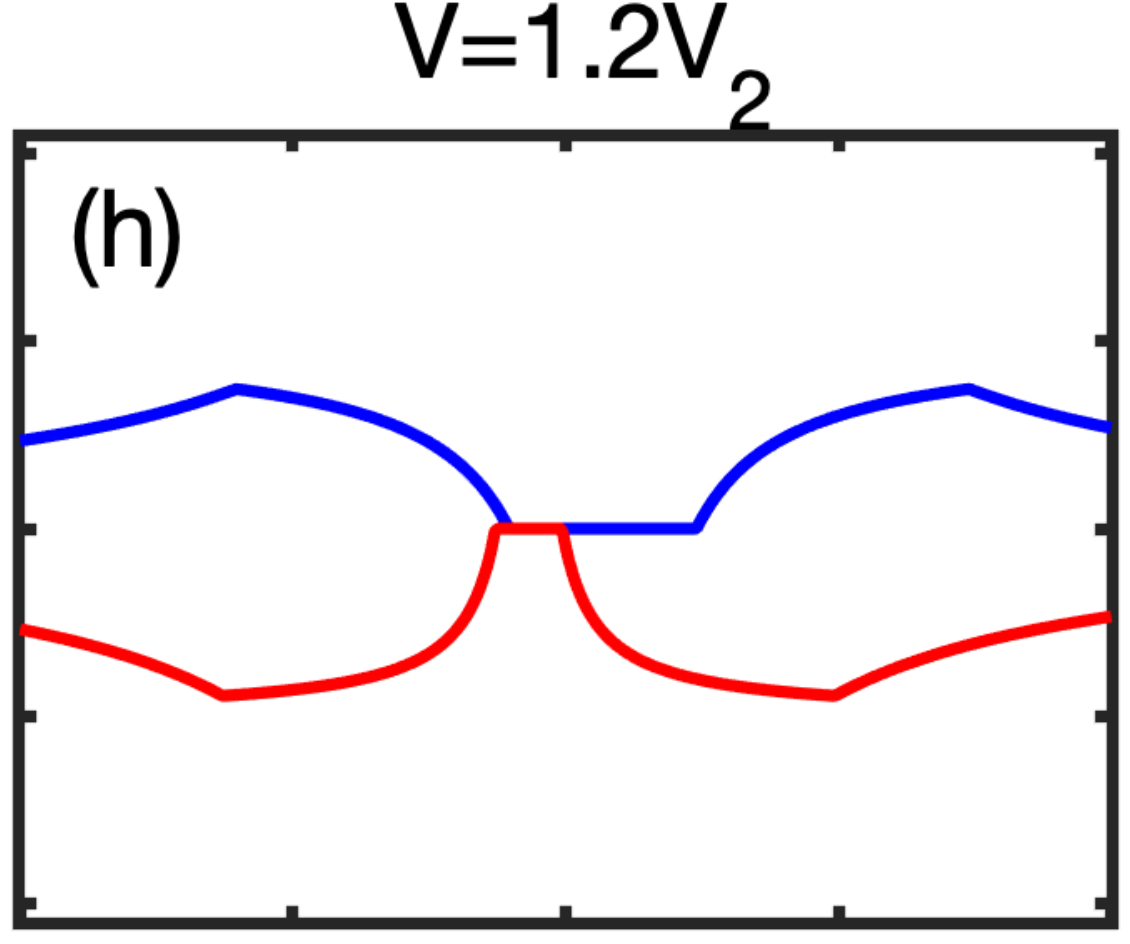}\\
\vspace{-0.05cm}
\hspace{-0.4cm}\includegraphics[height=3.35cm, width=4.7cm ]{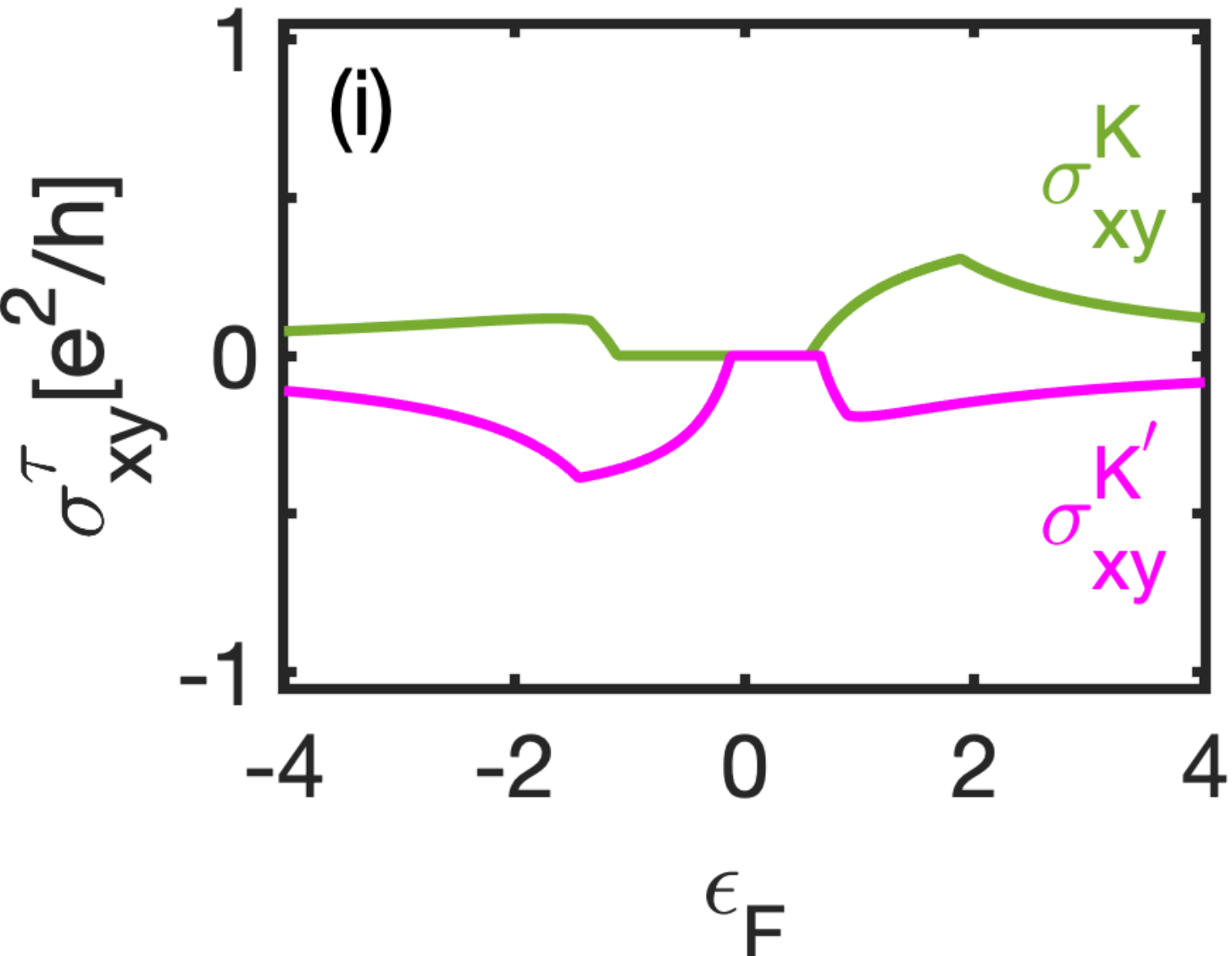}
\hspace{-0.15cm}\includegraphics[height=3.3cm, width=3.6cm ]{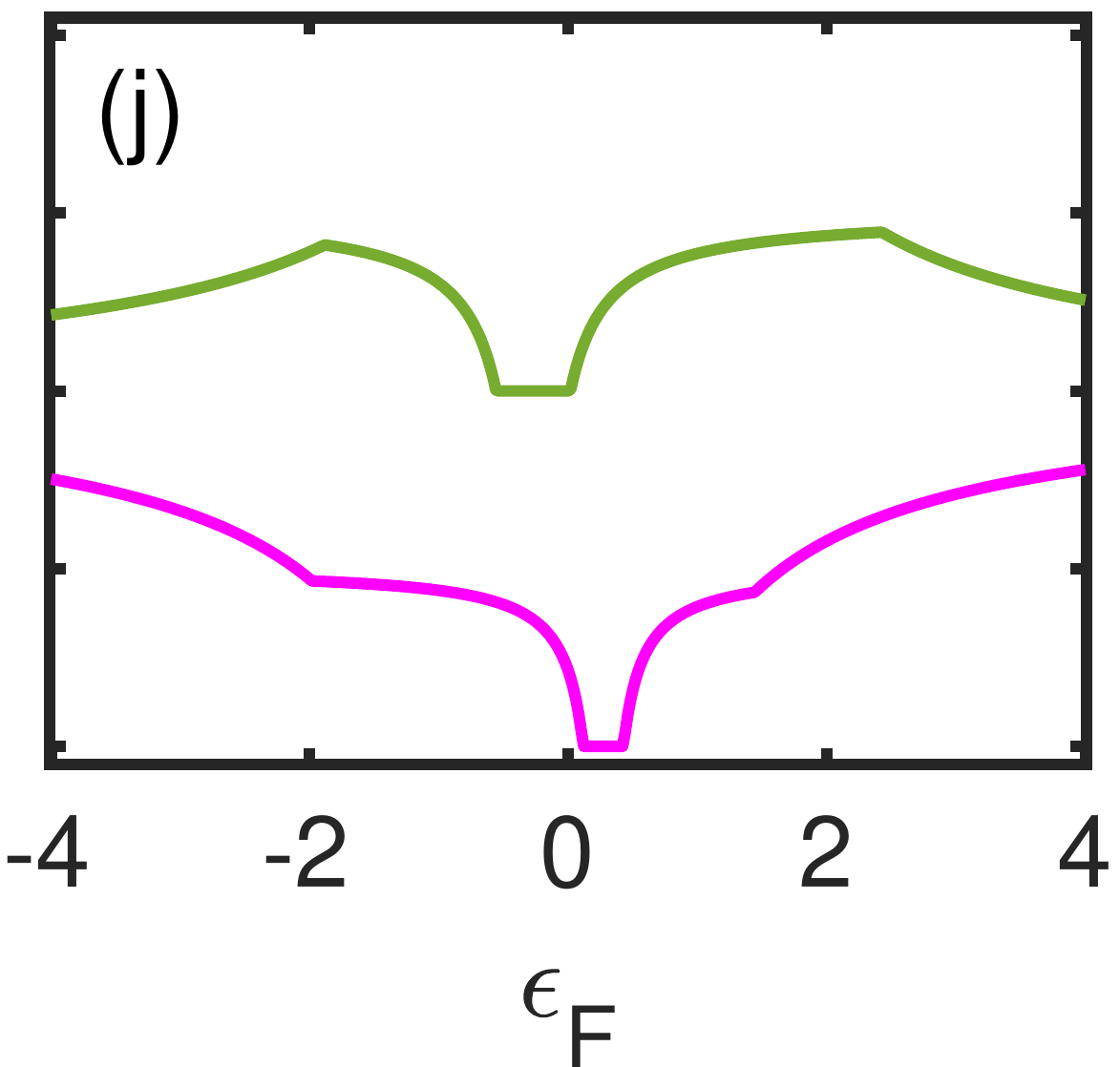}
\hspace{-0.14cm}\includegraphics[height=3.3cm, width=3.6cm ]{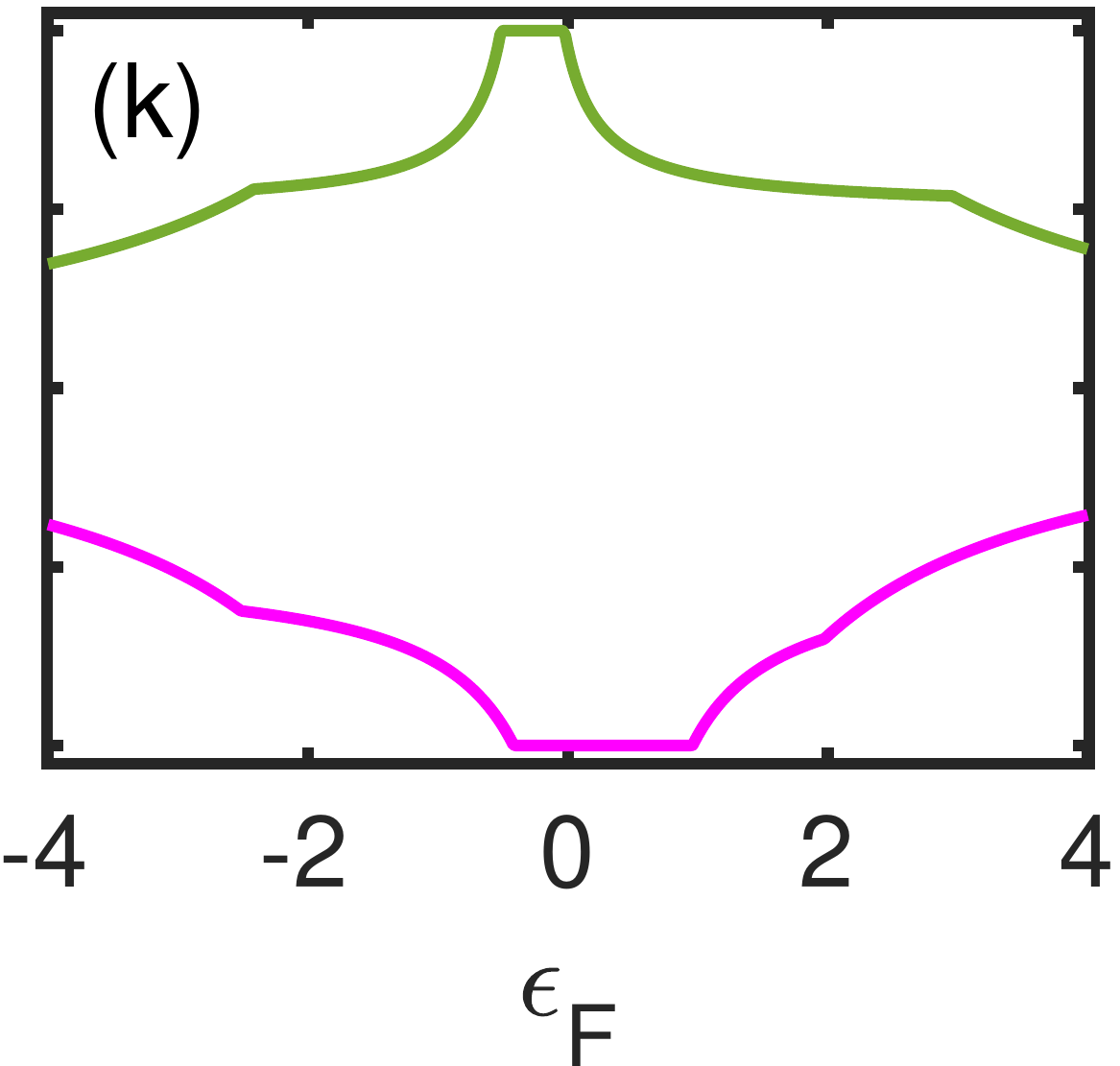}
\vspace*{-0.15cm} \caption{(Colour online) Evolution of the band structure of monolayer Pt$_2$HgSe$_3$ [(a)-(e)] with staggered exchange and Zeeman fields and increasing sublattice potential $V$. The parameters are $m_b = 0.5 \lambda_{so}$, $m_t = 0.05 \lambda_{so}$, $\lambda_{so} = 81.2$meV, and $\lambda_R = 0$. In (a) the system is a QSH insulator for $0 < V < V_1$ where $V_1 = \lambda_{so} - m_{-}$. After the gap closing and reopening of the spin-up channel at the $K^{\prime}$ valley for $V = V_1$ [shown in (b)], the system enters a VPM phase, shown in (c). After the second gap closing and reopening of the spin-down channel [shown in (d) at the $K$ valley for $V = V_2$ where $V_2 = \lambda_{so} + m_{-}$, the system becomes QVH insulator shown in (e). (f-h) Hall conductivities for the spin-up and spin-down channels, $\sigma_{xy}^{\uparrow}$ and $\sigma_{xy}^{\downarrow}$, respectively, as functions of $\epsilon_F = E_F / \lambda_{so}$. (i-k) Hall conductivities for each valley, $\sigma_{xy}^{K}$ and $\sigma_{xy}^{K^{\prime}}$ as functions of $\epsilon_F = E_F / \lambda_{so}$.   }
\label{fig:fig1}%
\end{figure*} 

\section{Chern numbers and topological phases}

We first consider the regime $m_b < \lambda_{so}$ and identify the topological phases as the sublattice potential $V$ varies. We calculate the Chern numbers, the Hall conductivities for each spin and valley, and use them to characterize these phases. The Rashba interaction is neglected ($\lambda_R = 0$) in this regime. The Berry curvature of a valence band is calculated from Eq.~(\ref{eq22}) together with Eqs.~(\ref{eqA4}) and (\ref{eqA5}) of the Appendix; it is given by
\begin{equation}
\Omega_{-}^{\tau s_z} ( k ) = \frac{\tau}{2} \frac{\hbar^2 v_F^2 M_{\tau s_z}}{\left( M_{\tau s_z}^2 + \epsilon_k^2 \right)^{3/2}}  .
\label{eq23}%
\end{equation} 
Integrating over the neighborhood of the $K$ or $K^{\prime}$ point, we obtain the Chern number as
\begin{equation}
\mathcal{C}_{\tau}^{s_z} = \frac{\tau}{2} \sgn ( m_{-} s_z + \lambda_{so} \tau s_z + V ) ,
\label{eq24}%
\end{equation} 
with $\sgn (x)$ the sign function, when the Fermi level $E_F$ is inside the insulating gap. A topological phase transition occurs when one of the four Dirac masses vanishes. 

When $E_F$ is in the conduction band we find analytically that the Hall conductivities for each spin component are given as
\begin{equation}
\sigma_{xy}^{s_z} = \pm \frac{e^2}{2 h} \sum_{\tau = \pm} \frac{\tau \left( m_{-} + \lambda_{so} \tau \pm V \right)}{\sqrt{\left( m_{-} + \lambda_{so} \tau \pm V \right)^2 + \hbar^2 v_F^2 k_F^2}}  ,
\label{eq25}%
\end{equation} 
where the $+(-)$ signs correspond to $s_z = \uparrow (\downarrow)$, respectively. We also find that the Hall conductivities for each valley take the form
\begin{eqnarray}
\hspace*{-0cm} 
\nonumber
\sigma_{xy}^{\tau} = \pm \frac{e^2}{2 h} \sum_{s_z = \pm} \frac{m_{-} s_z \pm \lambda_{so} s_z + V}{\sqrt{\left( m_{-} s_z \pm \lambda_ {so} s_z + V \right)^2 + \hbar^2 v_F^2 k_F^2}} , \\*
\label{eq26}%
\end{eqnarray}
where the $+ ( - )$ signs correspond to $\tau = K ( K^{\prime} )$ valleys.
 
\subsection{Case $V = 0$, $\lambda_R = 0$}
 
When $V=0$, we find that the Chern numbers for each spin channel are $\mathcal{C}_{\uparrow} = \mathcal{C}_{K}^{\uparrow} + \mathcal{C}_{K^{\prime}}^{\uparrow} = 1$ and $\mathcal{C}_{\downarrow} = \mathcal{C}_{K}^{\downarrow} + \mathcal{C}_{K^{\prime}}^{\downarrow} = -1$, leading to a Chern number $\mathcal{C} = 0$ and spin Chern number $\mathcal{C}_{s} = 2$. Thus, the gaps are topologically nontrivial for both spin channels, and the system is $\mathcal{T}$-symmetry-broken QSH insulator, i.e., it behaves as a QSH insulator with broken $\mathcal{T}$ symmetry. The Chern number contribution of each valley is found to be $\mathcal{C}_{K} = \mathcal{C}_{K}^{\uparrow} + \mathcal{C}_{K}^{\downarrow} = 0$ and $\mathcal{C}_{K^{\prime}} = \mathcal{C}_{K^{\prime}}^{\uparrow} + \mathcal{C}_{K^{\prime}}^{\downarrow} = 0$; the valley Chern number vanishes in this case $\mathcal{C}_{v} = 0$.

\subsection{Case $V \neq 0$, $\lambda_R = 0$}

As $V$ increases from zero, the system remains in the QSH phase, but the gap $\delta_{K^{\prime}}^{\uparrow}$ of the spin-up channel at $K^{\prime}$ valley shrinks, closes at $V = V_1$, and reopens for $V > V_1$, as shown in Figs.~3(a)-3(c). After reopening of the gap the system enters a VPM phase, first predicted in Ref.~\cite{eza12}, where part of the conduction (valence) band at $K ( K^{\prime} )$ valley is below (above) the Fermi level [see Fig.~3(c)]. In this phase, even though the gaps are open at the two valleys, the system becomes metallic; it also becomes valley polarized because electrons have moved from the $K$ valley to the $K^{\prime}$ valley. We can determine the range of $V$ in which the VPM phase exists by requiring $E_{K^{\prime} v}^{\uparrow} > 0$ and $E_{K c}^{\downarrow} <0$, where $E_{K^{\prime} v}^{\uparrow}$($E_{K c}^{\downarrow}$) are the spin up (spin down) valence (conduction) bands at $K^{\prime}$($K$). This gives
\begin{equation}
\vert V - \lambda_{so} \vert < m_{t}  \hspace{1cm}  \text{(VPM)}
\label{eq2300}%
\end{equation} 
Two remarks are in order here. First, if the Zeeman exchange were zero, i.e., $m_{+} = 0$, $E_F$ would be inside both gaps simultaneously. In this case, we confirmed that a spin-polarized QAH phase \cite{eza13,wu13} arises with $\mathcal{C}_{\uparrow} = 0$, $\mathcal{C}_{\downarrow} = -1$, $\mathcal{C}_s = 1$, and $\mathcal{C} = -1$; the system has only a spin-down edge current in this phase. Second, if the exchange field on the top Hg sublattice is taken to be zero, $m_t = 0$, we obtain marginal-VPM state (not shown here) where the conduction and valence bands touch the Fermi surface at the $K$ and $K^{\prime}$ points, respectively.
 
As $V$ increases further, the gap $\delta_{K}^{\downarrow}$ of the spin-down channel at the $K$ valley shrinks, closes at $V = V_2$, and reopens thereafter [see Figs.~3(d) and 3(e)]. After reopening of the gap, both spin channels become topologically trivial ($\mathcal{C}_{\uparrow} = \mathcal{C}_{\downarrow} = 0$), but the Chern number contribution of each valley is found to be $\mathcal{C}_{K} = 1$ and $\mathcal{C}_{K^{\prime}} = -1$; the system is a QVH insulator with valley Chern number $\mathcal{C}_v =2$.
\begin{figure}[t]
\vspace*{0.2cm}
\begin{center}
\hspace{0cm}\includegraphics[height=3.2cm, width=8.45cm ]{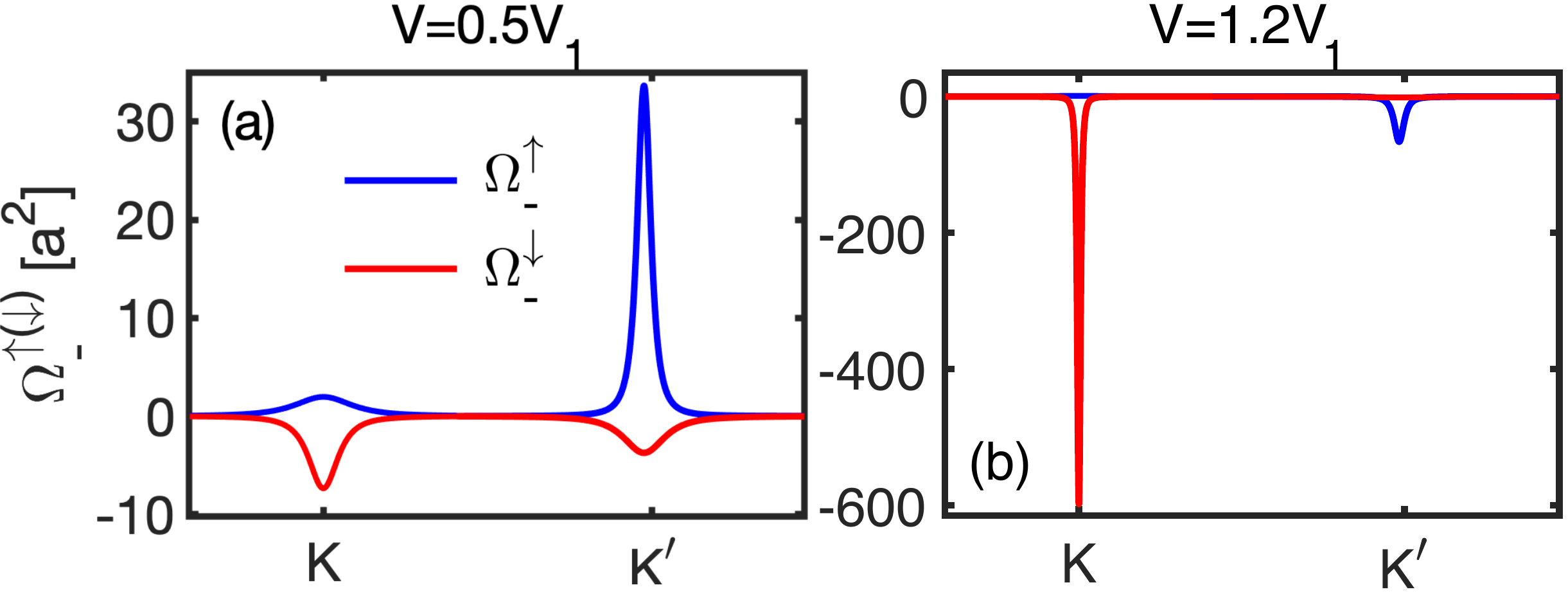}\\
\vspace{0.25cm}
\includegraphics[height=4.3cm, width=5.9cm ]{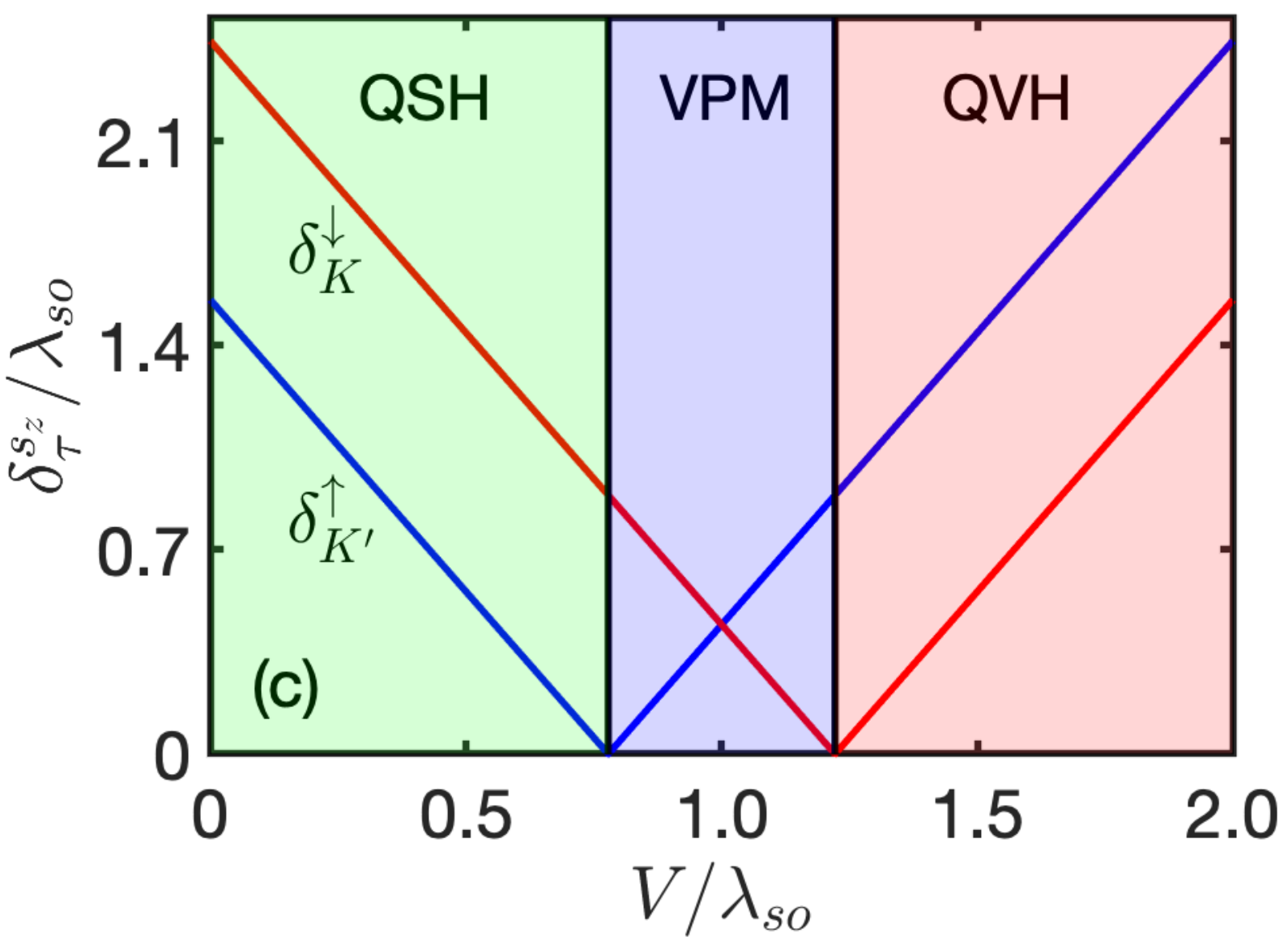}
\end{center}
\vspace*{-0.35cm} \caption{(Colour online) Berry curvature distribution of the spin-up and spin-down valence bands, $\Omega_{-}^{\uparrow}$ and $\Omega_{-}^{\downarrow}$, respectively, around the $K$ and $K^{\prime}$ points for (a) the QSH phase and (b) the VPM phase in units of $a^2$, where $a = 7.6$ \AA ~is the lattice constant. Other parameters are the same as in Fig.~3. (c) The phase diagram and the band gaps versus $V / \lambda_{so}$. }
\label{fig:fig1}%
\end{figure} 

Along the phase boundaries, $V = V_1$ and $V = V_2$, we observe that the system can host spin-polarized SDC semimetal states where one Dirac cone is massless and three Dirac cones are massive \cite{eza13}. Its creation needs broken $\mathcal{T}$ and $\mathcal{P}$ symmetries. The SDC state $u_{n s_z k}$ has the Berry's phase
\begin{equation}
\gamma_B = i \int_{0}^{2 \pi} d \varphi_k \langle u_{n s_z k} \vert \frac{\partial}{\partial \varphi_k} \vert u_{n s_z k}  \rangle = - \tau \pi  ,
\label{eq230}%
\end{equation} 
and the gap-closing at the valleys is tunable. It can also induce quantum Hall effect with no half-integer plateaux \cite{ezawa13}.
 
To illustrate further the topological transport properties of these phases, we show in Figs.~3(f)-3(h) the Hall conductivities for the spin-up and spin-down components, $\sigma_{xy}^{\uparrow}$ and $\sigma_{xy}^{\downarrow}$, respectively, as functions of the dimensionless Fermi energy $\epsilon_F = E_F / \lambda_{so}$. In Figs.~3(i)-3(k) we show the Hall conductivities for each valley, $\sigma_{xy}^{K}$ and $\sigma_{xy}^{K^{\prime}}$. Figure 3(f) exhibits the nontrivial topology of each spin channel in the QSH phase; the spin-up  and spin-down channels carry Hall conductances $e^2 / h$ and $- e^2 / h$, respectively, in the corresponding gaps. When $E_F$ is inside both gaps simultaneously the spin-Hall conductivity is quantized as $\sigma_{xy}^s = 2 e^2 / h$; an in-plane electric field drives the spin-up and spin-down electrons toward the opposite transverse edges of the sample, leading to a quantized spin Hall effect. This nontrivial topology can be traced to the distribution of the Berry curvatures shown in Fig.~4(a); the spin-up and spin-down channels have Berry curvatures of opposite sign, while their distribution is such that their integral over the $K$ and $K^{\prime}$ points yield $\mathcal{C}_{\uparrow} = 1$ and $\mathcal{C}_{\downarrow} = -1$. We also observe that $\sigma_{xy}^{K}$ and $\sigma_{xy}^{K^{\prime}}$ are vanishing in the corresponding gaps [see Fig.~3(i)] as a consequence of the vanishing integrated Berry curvature over the $K$ or $K^{\prime}$ point ($\mathcal{C}_{K} = \mathcal{C}_{K^{\prime}} = 0$). 

In the VPM state, which occurs for $V_1 < V < V_2$, the spin-up channel has zero Hall conductance while the spin-down channel has Hall conductance $- e^2 / h$. The distribution of the Berry curvature of the spin-up channel is such that its integral over $K$ and $K^{\prime}$ points vanishes, but the Berry curvature of the spin-down channel exhibits a sharp negative dip at the $K$ point whose integral gives $-1$ [see Fig.~4(b)]. In this state, the Fermi level does not lie inside both gaps simultaneously and the Chern number and spin Chern number are not well defined.

In the QVH phase, $\sigma_{xy}^{\uparrow} = 0$ and $\sigma_{xy}^{\downarrow} = 0$ in the corresponding gaps [see Fig.~3(h)], but $\sigma_{xy}^{K} = e^2 / h$ and $\sigma_{xy}^{K^{\prime}} = - e^2 / h$ [see Fig. 3(k)]. The valley Hall conductivity, $\sigma_{xy}^v = \sigma_{xy}^K - \sigma_{xy}^{K^{\prime}}$, which characterizes the accumulation of valley-resolved electrons to opposite sides of the sample is then quantized as $\sigma_{xy}^v = 2 e^2 / h$ when $E_F$ is inside both gaps at the same time. 

We provide a phase diagram and the band gaps as functions of $V / \lambda_{so}$ in Fig.~4(c). Blue and red lines correspond, respectively, to gaps at $K^{\prime}$ and $K$ of spin-up and spin-down channels. The QSH phase exists for $0 < V < V_1$, the VPM phase for $V_1 < V < V_2$, and the QVH phase for $V_2 < V < 2 \lambda_{so}$. Along the phase boundaries at $V = V_1 \simeq 0.78 \lambda_{so}$ or $V =V_2 \simeq 1.22 \lambda_{so}$, there exist spin-polarized SDC semimetal states where the gaps $\delta_{K^{\prime}}^{\uparrow}$ and $\delta_{K}^{\downarrow}$ close. 

In Fig.~5 we show the spin-Hall conductivity $\sigma_{xy}^s$ as a function of $\epsilon_F = E_F / \lambda_{so}$ for the same parameter values as in Fig.~3(f) and increasing temperature which is included through the Fermi function in Eq.~(\ref{eq22}). We do not consider electron-phonon interactions here. For the parameters used in this work, the plateau extends over $55$meV and is visible for temperatures up to $\sim 90$K. For comparison, in unmagnetized silicene the quantized spin Hall effect persists up to $\sim 10$ K \cite{vargiam14}.

\section{Valley-polarized QAH effect}

Below we explore the second regime, $m_b > \lambda_{so}$, and we show that the system is VP-QAH insulator for $0 < V < m_{+}$. For $V > m_{+}$ it transitions first to VPM and then to QVH  insulator. The VP-QAH phase combines valleytronics and topology, i.e., the properties of both the QAH phase and the QVH phase coexist in one material. Importantly, we also find that reversing the sign of the exchange interaction swaps the topology of each valley, indicating that the Chern number is coupled to the substrate's magnetization. This conclusion is promising for a magnetic manipulation of the VP-QAH effect. 

We compute the Chern numbers from Eq.~(\ref{eq21}), where the Berry curvature is expressed as
\begin{equation}
\Omega_{n s} ( k ) = i \langle \nabla_{\mathbf{k}} u_{n s k} \vert \times \vert \nabla_{\mathbf{k}} u_{n s k} \rangle  ,
\label{eq27}%
\end{equation} 
with $u_{n s k}$ denoting the Bloch state for a band labeled by $n$ and $s$; they are given in Eqs.~(\ref{eq14}) and (\ref{eq15}). This expression for the Berry curvature is computationally more convenient here. Using the polar coordinate system, we find that the Berry curvature for a valence band ($n = -$) with $s = \pm$ is expressed as
\begin{equation}
\Omega_{- s} ( k ) = \frac{1}{k} \frac{\partial}{\partial k} Q_{-s} ( k )  ,
\label{eq28}%
\end{equation} 
\begin{figure}[t]
\vspace*{0.2cm}
\begin{center}
\hspace{-0.2cm}\includegraphics[height=5.5cm, width=7.4cm ]{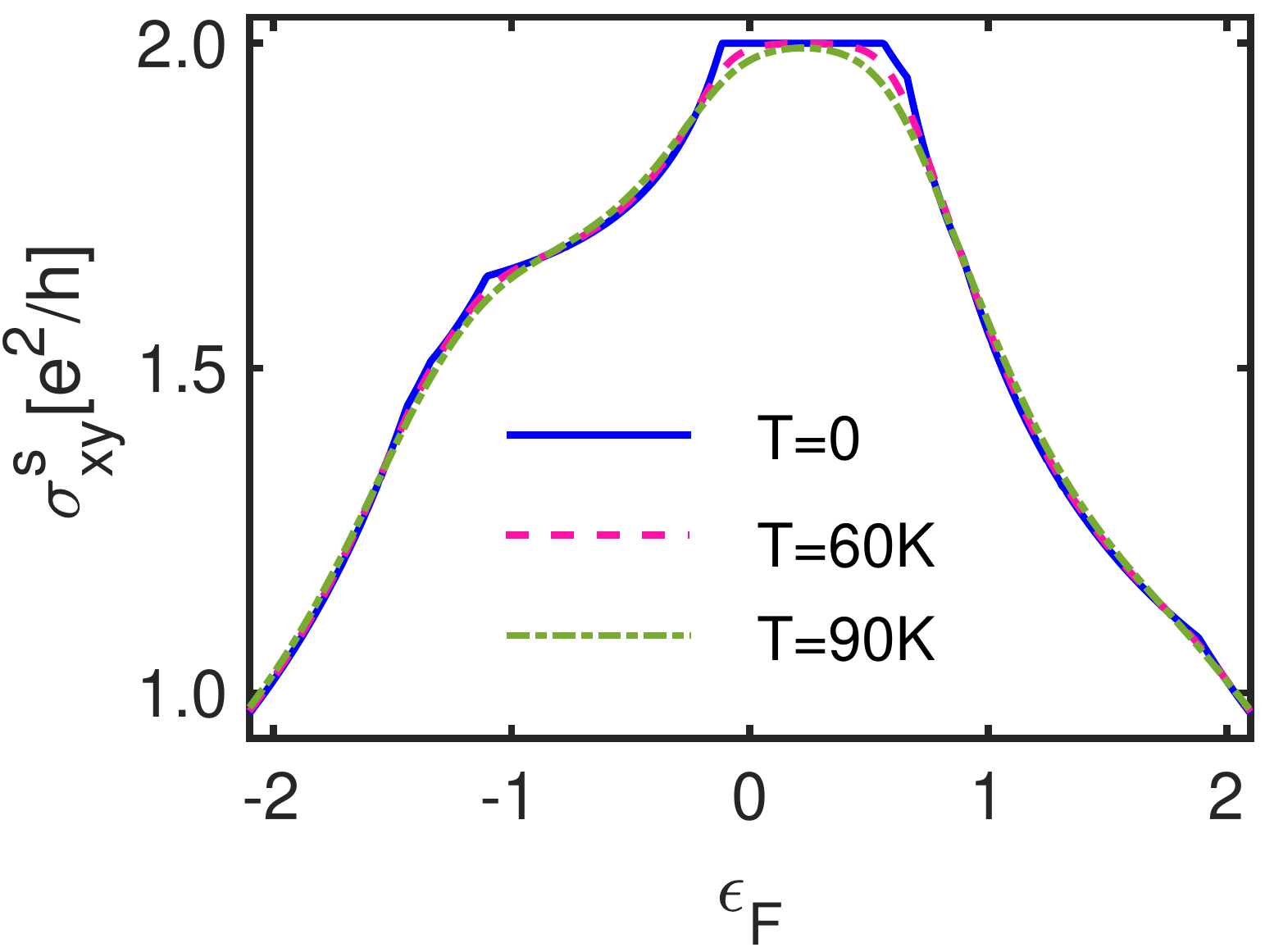}
\end{center}
\vspace*{-0.3cm} \caption{(Colour online) Spin-Hall conductivity for the same parameters as in Fig.~3(f) at different temperatures.   }
\label{fig:fig1}%
\end{figure} 
with
\begin{align}
\nonumber \hspace*{-0.0cm}
Q_{- s}^{K} ( k ) = N^2 \Big[ \Big( \frac{E_{- s k} - \gamma_2^{\downarrow}}{\epsilon_k}\Big)^2 \hspace{-0.1cm}- \hspace{-0.1cm}2 \Big( \frac{\epsilon_k e^{i \varphi_k} \eta}{E_{- s k} - \gamma_{1}^{\uparrow}} \Big)^2 \hspace{-0.08cm} - \hspace{-0.08cm}e^{2 i \varphi_k} \eta^2  \Big]  , \\*
\label{eq29}%
\end{align}
for the $K$ valley and
\begin{equation} 
\hspace*{-0.2cm} 
Q_{- s}^{K^{\prime}} ( k ) = - \tilde{N}^2 \epsilon_k^2 \Big[ \Big( \frac{\tilde{\eta}}{E_{- s k} - \gamma_{2}^{\uparrow}}\Big)^2 + \Big( \frac{1}{E_{- s k} - \gamma_1^{\downarrow}}\Big)^2  \Big], 
\label{eq30}%
\end{equation} 
for the $K^{\prime}$ valley. In the following, we will omit the $-$ sign for a valence band and write $Q_{K s}$ and $Q_{K^{\prime} s}$ for brevity. Substituting 
$\Omega_{- s} ( k )$ from Eq.~(\ref{eq28}) into Eq.~(\ref{eq21}) we find 
\begin{equation}
\mathcal{C}_{K ( K^{\prime} ) s} = \left[ Q_{K ( K^{\prime} ) s} ( \infty ) - Q_{K ( K^{\prime} ) s} ( 0 ) \right]  .
\label{eq31}%
\end{equation}
\begin{figure*}[t]
\vspace*{0.2cm}
\begin{center}
\includegraphics[height=7.8cm, width=17.7cm ]{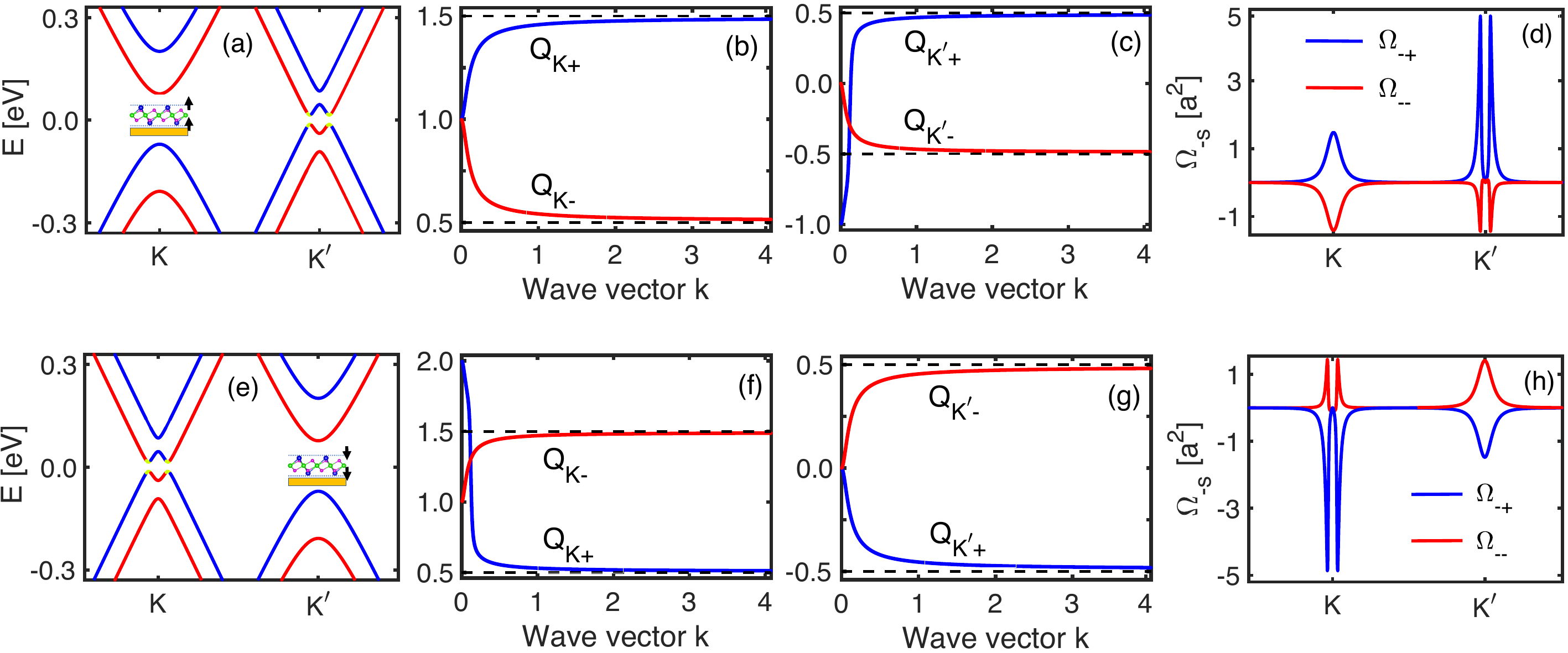}
\end{center}
\vspace*{-0.15cm} \caption{(Colour online) (a) Band structure near the $K$ and $K^{\prime}$ valleys for $m_b = 1.5 \lambda_{so}$, $m_t = 0.05 \lambda_{so}$, $\lambda_{R} = 0.15 \lambda_{so}$, and $V = 0$. (b)-(c) Calculated $Q_{Ks}$ and $Q_{K^{\prime}s}$ as functions of the wave vector $k$. (d) Berry curvatures of the $s = +$ and $s=-$ valence bands, $\Omega_{- +}$ and $\Omega_{- -}$, near the $K$ and $K^{\prime}$ valleys. (e)-(h) The same as in (a)-(d) but with opposite exchange fields $m_b = - 1.5 \lambda_{so}$ and $m_t = - 0.05 \lambda_{so}$.    }
\label{fig:fig1}%
\end{figure*} 

\subsection{Case $V = 0$, $\lambda_R \neq 0$}

In Fig.~6(a) we show the band structure around the $K$ and $K^{\prime}$ valleys for $m_b = 1.5 \lambda_{so}$, $m_t = 0.05 \lambda_{so}$, $\lambda_R = 0.15 \lambda_{so}$, and $V = 0$. Numerical calculation shows that $Q_{K +} ( \infty ) \simeq 1.5$, $Q_{K -} ( \infty ) \simeq 0.5$, and $Q_{K +} ( 0 ) = Q_{K -} ( 0 ) = 1$, as shown in Fig.~6(b). Then Eq.~(\ref{eq31}) gives
\begin{equation}
\mathcal{C}_{K \pm } \simeq \pm 0.5  .
\label{eq32}%
\end{equation} 
The individual Chern numbers $\mathcal{C}_{K \pm }$ are not exactly quantized but depend numerically on the value of $\lambda_R$. However, the two contributions always sum up to $0$ so the Chern number of valley $K$ is $\mathcal{C}_K = 0$. For the $K^{\prime}$ valley we find $Q_{K^{\prime} \pm} ( \infty ) \simeq \pm 0.5$, $Q_{K^{\prime} +} ( 0 ) = - 1$, and $Q_{K^{\prime} -} ( 0 ) = 0$, as shown in Fig.~6(c). Therefore,
\begin{equation}
\mathcal{C}_{K^{\prime} + } \simeq 1.5  ,\qquad
\mathcal{C}_{K^{\prime} - } \simeq - 0.5  ,
\label{eq34}%
\end{equation} 
and the Chern number of valley $K^{\prime}$ is $\mathcal{C}_{K^{\prime} } = 1$. It follows that the gap at $K ( K^{\prime} )$ is topologically trivial (nontrivial) with total Chern number $\mathcal{C} = 1$. In addition, the different Chern numbers of valleys $K$ and $K^{\prime}$ give rise to nonzero valley Chern number $\mathcal{C}_v = - 1$, indicating the existence of VP-QAH effect with a single edge mode. The topologically different responses of valleys $K$ and $K^{\prime}$ arise from the strong intrinsic SOC and the fact that the gaps are valley-dependent.

In Fig.~6(d) we show the distribution of the Berry curvature around the $K$ and $K^{\prime}$ valleys for each band. We notice that the Berry curvatures of the two bands at the $K$ valley, $\Omega_{- +}$ and $\Omega_{- -}$, are exactly equal and opposite such that their $k$-space integrals yield $\mathcal{C}_{K +} = - \mathcal{C}_{K -} \simeq 0.5$ leading to $\mathcal{C}_K = 0$. The Berry curvatures of the two bands at the $K^{\prime}$ valley are obviously different from those near the $K$ valley with unequal Chern numbers, and exhibit sharp peaks and dips. Their distribution is such that their integrals yield $\mathcal{C}_{K^{\prime}} = 1$.

In Fig.~6(e) we show the band structure with the sign of the exchange interaction reversed. In this case, we find $Q_{K +} ( \infty ) \simeq 0.5$, $Q_{K -} ( \infty ) \simeq 1.5$, $Q_{K +} ( 0 ) = 2$, and $Q_{K -} ( 0 ) = 1$, as shown in Fig.~6(f). It follows that
\begin{equation}
\mathcal{C}_{K + } \simeq - 1.5  ,\qquad
\mathcal{C}_{K - } \simeq  0.5  ,
\label{eq36}%
\end{equation}
and the Chern number of valley $K$ is $\mathcal{C}_K = -1$. For the $K^{\prime}$ valley [see Fig.~6(g)] we find $Q_{K^{\prime} \pm} ( \infty ) \simeq \mp 0.5$, and $Q_{K^{\prime} +} ( 0 ) = Q_{K^{\prime} -} ( 0 ) = 0$, which yield
 \begin{equation}
\mathcal{C}_{K^{\prime} \pm } \simeq \mp 0.5  ,
\label{eq37}%
\end{equation} 
and the Chern number of valley $K^{\prime}$ is $\mathcal{C}_{K^{\prime}} = 0$. Therefore, the gap at $K ( K^{\prime} )$ is topologically nontrivial (trivial) with Chern number $\mathcal{C} = - 1$ and valley Chern number $\mathcal{C}_v = - 1$; the topology is swapped between the two valleys, and the opposite sign of the Chern number indicates that the chirality of the edge state is reversed.

The swapping of the VP-QAH phase between the two valleys is also reflected in the Berry curvatures around the $K$ and $K^{\prime}$ points, as shown in Fig.~6(h). We observe that they have opposite signs from those in Fig.~6(d), and they have also been switched between the two valleys. 
\begin{figure}[t]
\vspace*{0.2cm}
\begin{center}
\includegraphics[height=3.8cm, width=8.6cm ]{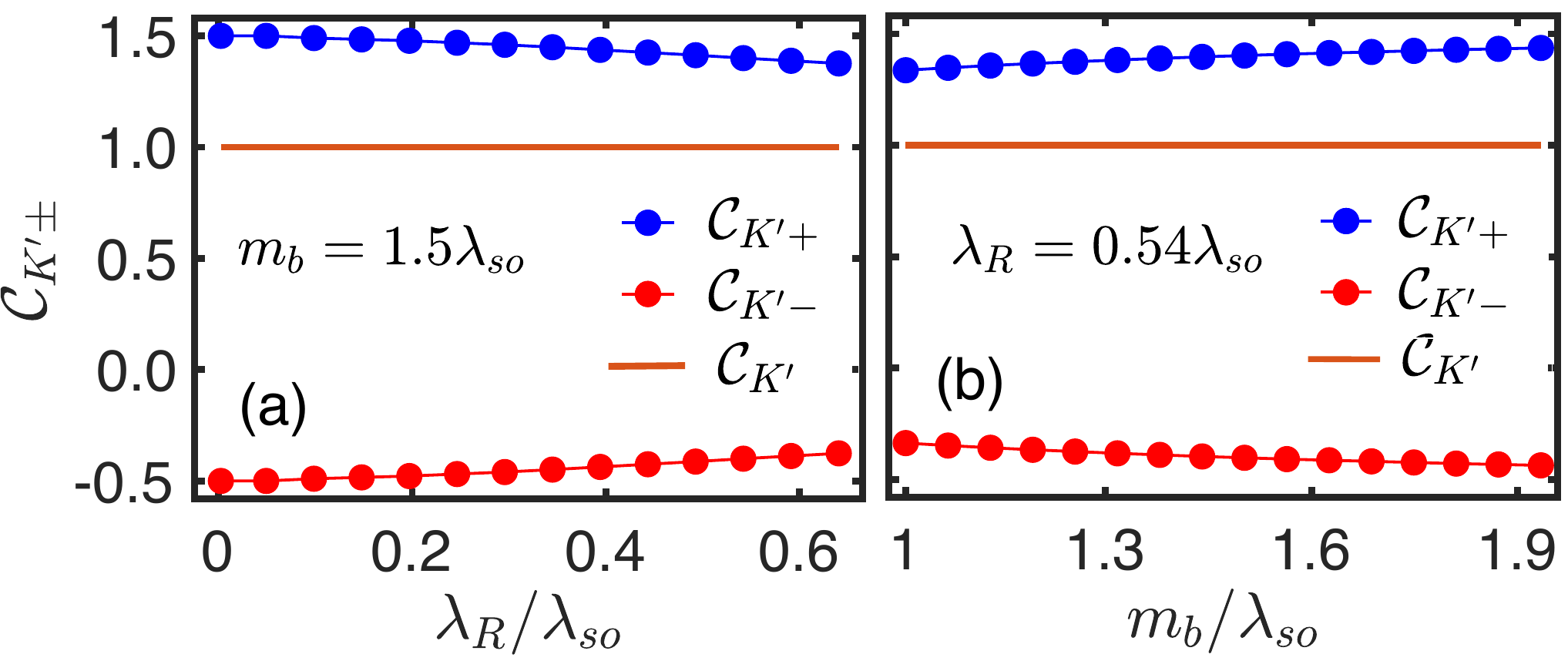}
\end{center}
\vspace*{-0.3cm} \caption{(Colour online) (a) Chern numbers for each valence band ($s = \pm 1$) in the topological gap at the $K^{\prime}$ valley as a function of $\lambda_{R}$ for fixed $m_b = 1.5 \lambda_{so}$. Their sum is always $\mathcal{C}_{K^{\prime}} = \mathcal{C}_{K^{\prime}+} + \mathcal{C}_{K^{\prime}-} = 1$ (solid, brown line). (b) Chern numbers for each valence band ($s = \pm 1$) at the $K^{\prime}$ valley as a function of the exchange field in the bottom sublattice $m_b$ for fixed Rashba SOC $\lambda_R = 0.54 \lambda_{so}$. Their sum is again $\mathcal{C}_{K^{\prime}} = 1$.   }
\label{fig:fig1}%
\end{figure} 

From the above results, the Hall conductivity for $V = 0$ can be expressed as
\begin{equation}
  \sigma_{xy}=\left\{
  \begin{array}{@{}ll@{}}
    e^2 / h \sgn ( m_b ), &  \vert m_b \vert > \lambda_{so} \\
    0, & \vert m_b \vert < \lambda_{so}
  \end{array}\right. ,
\label{eq38}%
\end{equation} 
where the contribution to the conductivity comes only from the $K^{\prime}$ valley when $m_b  > \lambda_{so}$, whereas for $m_b  < - \lambda_{so}$ the contribution comes only from the $K$ valley. The valley-Hall conductivity, $\sigma_{xy}^v = \sigma_{xy}^K - \sigma_{xy}^{K^{\prime}}$, which characterizes the accumulation of valley-resolved electrons to opposite sides of the sample is then given by
\begin{equation}
  \sigma_{xy}^v=\left\{
  \begin{array}{@{}ll@{}}
    - e^2 / h, &  \vert m_b \vert > \lambda_{so} \\
    0, & \vert m_b \vert < \lambda_{so}
  \end{array}\right. ,
\label{eq39}%
\end{equation} 
\begin{figure*}[t]
\vspace*{0.2cm}
\begin{center}
\hspace*{-0cm}\includegraphics[height=3.3cm, width=17.9cm ]{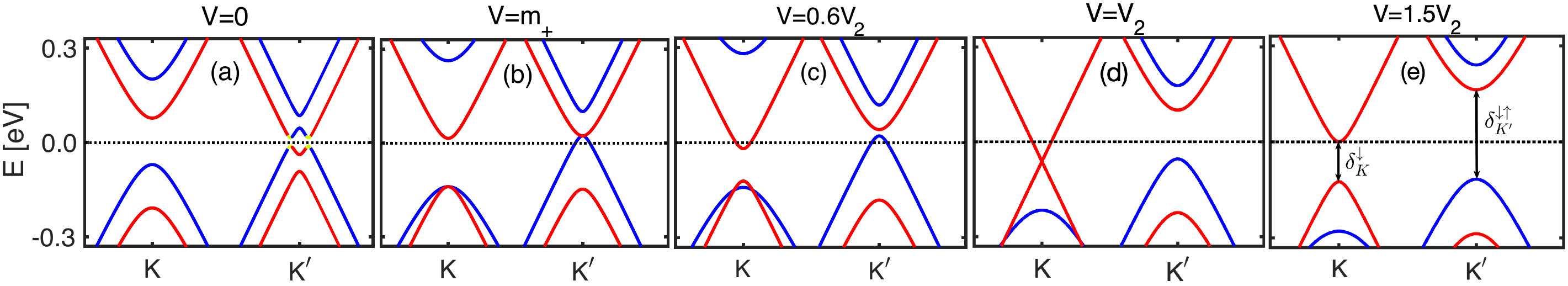}
\end{center}
\vspace*{-0.2cm} \caption{(Colour online) (a)-(e) Evolution of the band structure with increasing $V$ for $\lambda_R = 0.15 \lambda_{so}$. The other parameters are the same as in Fig.~6(a). For $0 < V < m_{+}$ the system is VP-QAH insulator whereas for $m_{+} < V < V_2$ it is VPM. At $V = V_2$ the gap $\delta_K^{\downarrow}$ of the spin-down channel at the $K$ valley closes and reopens and the system becomes QVH insulator. }
\label{fig:fig1}%
\end{figure*} 

The Chern number contribution of each valence band and the total Chern number are shown in Fig.~7 as functions of Rashba SOC $\lambda_R$ and exchange field $m_b$ for the $K^{\prime}$ valley. In Fig.~7(a) the exchange field is fixed at $m_b = 1.5 \lambda_{so}$ and in Fig.~7(b) the Rashba SOC is fixed at $\lambda_R = 0.54 \lambda_{so}$. In Fig.~7(a) we notice that in the limit $\lambda_R \rightarrow 0$, the Chern number for the valence band with $s = -1$ is negatively half quantized, i.e., $\mathcal{C}_{K^{\prime}-} = - 0.5$, and that for the valence band with $s = +1$ is one and half quantized, i.e., $\mathcal{C}_{K^{\prime}+} = 1.5$. For increasing values of $\lambda_R$, their absolute values, $\vert \mathcal{C}_{K^{\prime}+} \vert$ and $\vert \mathcal{C}_{K^{\prime}-} \vert$, reduce but their sum is always one, i.e., $\mathcal{C}_{K^{\prime}} = \mathcal{C}_{K^{\prime}+} + \mathcal{C}_{K^{\prime}-} = 1$ (solid, brown line). In Fig.~7(b) one observes that for $m_b \rightarrow 2 \lambda_{so}$ the contributions are $\mathcal{C}_{K^{\prime}-} =  - 0.5$ and $\mathcal{C}_{K^{\prime}+} =  1.5$. For smaller values of $m_b$ their absolute values reduce, but their sum is again always one, $\mathcal{C}_{K^{\prime}} = 1$.  
\begin{figure}[b]
\vspace*{0.2cm}
\begin{center}
\hspace{-0.2cm}\includegraphics[height=5.5cm, width=7.4cm ]{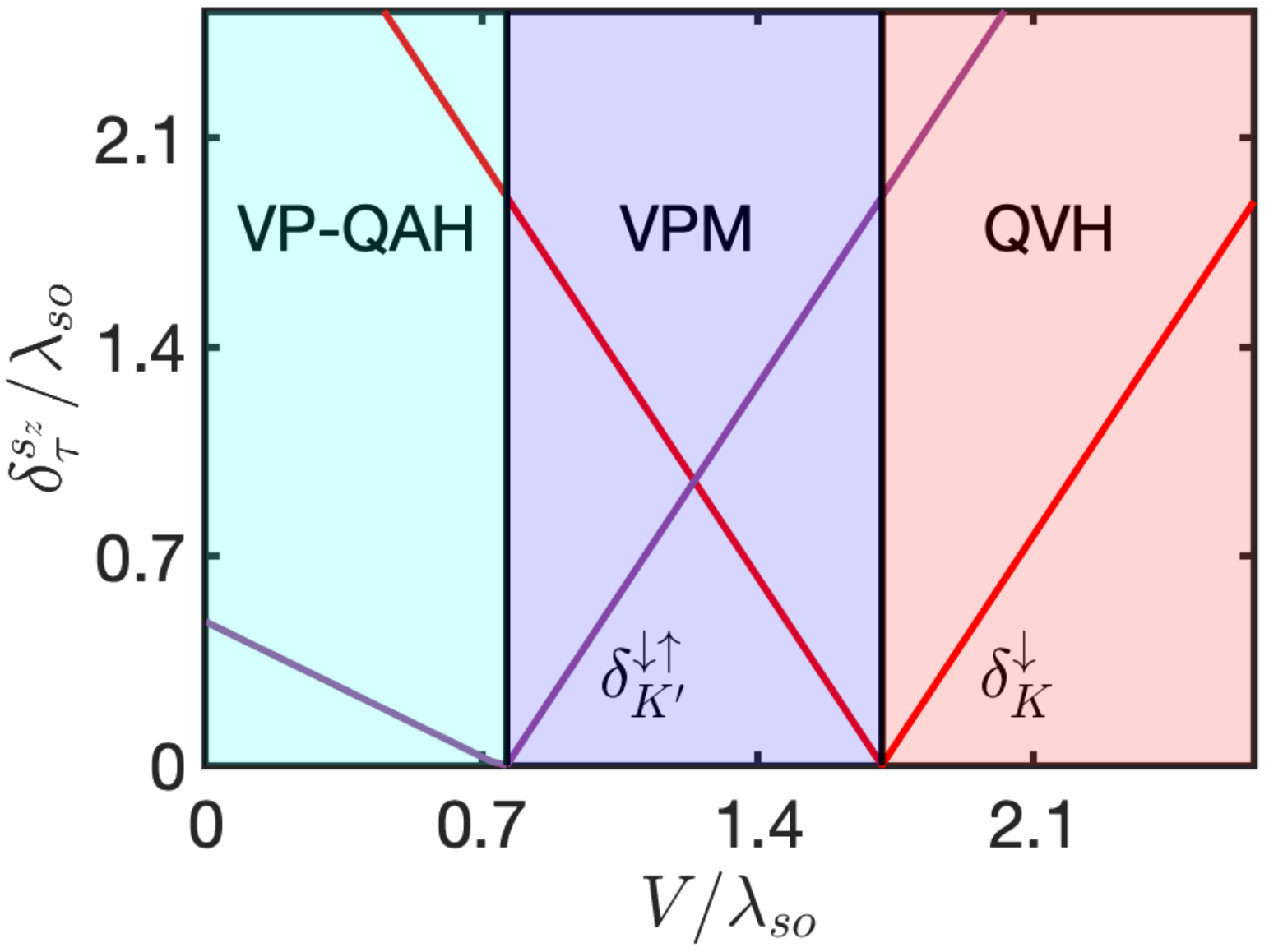}
\end{center}
\vspace*{-0.3cm} \caption{(Colour online) (a) The phase diagram and the band gaps as functions of $V / \lambda_{so}$. In the VP-QAH phase, which exist for $0 < V < m_{+}$, the Chern number and valley Chern number are $\mathcal{C} = 1$ and $\mathcal{C}_v = - 1$. In the QVH phase, which exists for $V > V_2$, $\mathcal{C} = 0$ and $\mathcal{C}_v = 2$. Violet and red lines correspond to band gaps at the $K^{\prime}$ and $K$ valleys.  }
\label{fig:fig1}%
\end{figure}

\subsection{Case $V \neq 0$, $\lambda_R \neq 0$}

As $V$ increases from zero [see Fig.~8(a)] up to $V = m_{+}$, the gap $\delta_K^{\downarrow \uparrow}$ remains open with $\mathcal{C} = 0$ and the gap $\delta_{K^{\prime}}^{\downarrow \uparrow}$ remains topological with $\mathcal{C}_{K^{\prime}} = 1$ suggesting that the VP-QAH phase is robust against the staggered sublattice potential, i.e., for  $0 < V < m_{+}$. The gap $\delta_{K^{\prime}}^{\downarrow \uparrow}$ closes at $V = m_{+}$, and reopens for $V > m_{+}$ as shown in Figs. 8(b) and 8(c), and the system undergoes a phase transition from a VP-QAH phase to VPM phase (see also Fig.~9). As $V$ increases further, the gap $\delta_K^{\downarrow}$of the spin-down channel decreases, closes at $V = V_2 = \lambda_{so} + m_{-}$, and reopens for $V > V_2$, as shown in Figs.~8(d) and 8(e). After reopening of the gap, the Chern number contribution of each valley is $\mathcal{C}_K = 1$ and $\mathcal{C}_{K^{\prime}} = - 1$; the system becomes QVH insulator with valley Chern number $\mathcal{C}_v = 2$. 

In Fig.~9 we show the topological phases and band gaps as functions of $V / \lambda_{so}$. The VP-QAH phase exists for $0 < V < m_{+}$ ($m_{+} \simeq 0.75 \lambda_{so}$). Along the phase boundary at $V = m_{+}$, there exists a single-valley topological metal (TM) state; the bands touch parabolically in this state, see Fig.~8(b). At the phase boundary at $V = V_2 = \lambda_{so} + m_{-} \simeq 1.7 \lambda_{so}$ there exists a SDC state. The violet line corresponds to the topological gap $\delta_{K^{\prime}}^{\downarrow \uparrow}$ at the $K^{\prime}$ valley. It decreases for $0< V < m_{+}$, becomes zero at $V = m_{+}$, and increases thereafter. The gap $\delta_K^{\downarrow}$of the spin-down channel (red line) closes and reopens at $V = V_2$.

\section{Summary and conclusions}

We investigated topological phases in monolayer jacutingaite with exchange fields, staggered sublattice potential, and Rashba SOC. Jacutingaite is a naturally occuring layered mineral and its monolayer displays the KM physics but at a much higher energy scale than other 2D honeycomb materials like graphene, silicene, germanene, and stanene. Our analysis is based on topological invariants and reveals that the system exhibits QSH, VPM, VP-QAH, and QVH phases, that are tunable by the exchange fields and/or an external electric field. We also demonstrate that these phases can be characterized and distinguished by the spin- and valley-Hall conductivities.

In particular, for $m_b < \lambda_{so}$ and $0 < V < V_1$, we find that the system exhibits a quantized spin-Hall effect with spin-Hall conductivity $\sigma_{xy}^s = 2 e^2 / h$, despite the broken $\mathcal{T}$ symmetry. For $V_1 < V < V_2$ it is VPM, and for $V > V_2$ it exhibits a quantized valley-Hall effect with valley-Hall conductivity $\sigma_{xy}^v = 2 e^2 / h$. Along the phase boundaries we find that SDC states emerge. 

For $m_b > \lambda_{so}$ and finite Rashba SOC, our analysis shows that the system is VP-QAH insulator with quantized Chern number $\mathcal{C} = 1$ and quantized valley Chern number $\mathcal{C}_v = -1$, i.e., it exhibits simultaneously the properties of the QAH effect and the QVH effect. The VP-QAH phase is robust against the staggered sublattice potential and exists for $0 < V < m_{+}$. Further, we show that reversing the sign of the exchange interactions switches the topological properties between the two valleys; for $m_b < - \lambda_{so}$ the valley polarization is switched and the Chern number is $\mathcal{C} = - 1$. The apparent coupling of the Chern number to the substrate's magnetization can lead to the magnetic manipulation of the VP-QAH phase. As $V$ increases further, the system exhibits VPM and QVH phases. The intriguing possibility of realizing the VP-QAH insulator can lead to topological valleytronics applications, e.g., the topological valley field-effect transistor \cite{qian14}.

We remark that valley-contrasted topological thermoelectric transport in the system described here can be an interesting research direction as well. Topological valley-dependent anomalous thermoelectric transport in bilayer transition metal dichalcogenides has been recently investigated in Ref.~\cite{vargiam20}.

In experiments the exchange fields can be generated using a bulk magnetic insulator, for example, EuO or EuS. However, electrical switching and tuning the substrate magnetization is challenging to achieve in traditionall 3D magnets. Currently, there is an intense focus on 2D magnetic crystals which allow electrical, magnetic, and optical control \cite{gong19,zhong17,seyler18}, which makes them engineerable and integrable into vdW heterostructures. Recently, it was experimentally demonstrated that 2D chromium sulfide bromide (CrSBr) (an air-stable vdW semiconductor with band gap $\sim 1.5$eV and interlayer antiferromagnetic ordering up to relatively high N\'{e}el temperature of $\sim 132$K \cite{wang20}), provides graphene with strong exchange interaction and considerable spin-splitting of $\sim 20$meV \cite{kaver21}. Monolayer CrSBr is a ferromagnet with a predicted Curie temperature of $\sim 168$K (potentially even higher) and gate-tunable magnetization \cite{wang20}. These properties make CrSBr a promising 2D magnet for the realization of the VP-QAH effect which should be stable up to high temperature. This temperature stability is an important parameter for realistic applications of the VP-QAH effect. 

Besides CrSBr, other magnetic crystals are possible candidates, for example CrI$_3$ \cite{zhong17,seyler18}, which shows Ising-type magnetization down to mono- and bi-layer limits and allows electrical control of the magnetization \cite{shan18}. It is also promising as substrate material to realize these topological properties.

Finally, we note that the VP-QAH phase was also found in the vdW heterostructure Pt$_2$HgSe$_3$/CrI$_3$ investigated in Ref.~\cite{liu20}. However, in this work the authors focus only in the regime $m_b > \lambda_{so}$ with $m_t = 0$, whereas the effect of staggered sublattice potential was not considered.

\begin{appendix}

\section{Invariance of $m_{-} \sigma_z s_z$ under $\mathcal{PT}$ symmetry}

Under inversion symmetry, $\textbf{r} \rightarrow - \textbf{r}$, $\textbf{p} \rightarrow - \textbf{p}$, and $\textbf{s} \rightarrow \textbf{s}$. Inversion is represented by a unitary operator,
\begin{equation}
\mathcal{P} = \sigma_x \tau_x ,
\label{eqA12}%
\end{equation}
where $\tau_x$ switches the valleys. Under time reversal, $\textbf{r} \rightarrow  \textbf{r}$, $\textbf{p} \rightarrow - \textbf{p}$, and $\textbf{s} \rightarrow - \textbf{s}$. The time reversal symmetry for spin $1 / 2$ particles is represented by the operator,
\begin{equation}
\mathcal{T} = \tau_x i s_y K_c ,
\label{eqA22}%
\end{equation}
where $i s_y$ produces reversal of the electron spin, and $K_c$ is the complex conjugation. For spin $1 / 2$ electrons, $\mathcal{T}$ has the property $\mathcal{T}^2 = - 1$. The preservation of $m_{-} \sigma_z s_z$ under the operator product of inversion and time reversal, $\mathcal{PT}$, is shown readily 
using the standard anticommutation relations $\lbrace \sigma_i, \sigma_j \rbrace = 2 \delta_{ij}$:
\begin{eqnarray}
\nonumber \hspace*{-0.13in} \mathcal{PT} (m_{-} \sigma_z s_z) (\mathcal{PT})^{-1} = \sigma_x i s_y ( m_{-} \sigma_z s_z ) \sigma_x i s_y^{\ast}
\\* && \nonumber \hspace*{-1.39in} = -\sigma_x  (m_{-} \sigma_z s_z) s_y \sigma_x s_y    
\\* && \nonumber \hspace*{-1.39in} = (m_{-} \sigma_z s_z) \sigma_x^2 s_y^2   
\\* && \hspace*{-1.39in} = m_{-} \sigma_z s_z   .
\label{eqA33}%
\end{eqnarray} 

\vspace{-0cm}
\section{Band gaps between opposite spin bands}

The gaps between spin-down conduction band and spin-up valence band at $K$ and $K^{\prime}$ valleys, $\delta_{K}^{\downarrow \uparrow}$ and $\delta_{K^{\prime}}^{\downarrow \uparrow}$, respectively, are given by
\begin{eqnarray} 
&&\hspace{-0.5cm}\delta_{K}^{\downarrow \uparrow} = - 2 m_{+} +\vert \lambda_{so} + m_{-} - V \vert + \vert \lambda_{so} + m_{-} + V \vert,\\
\label{eqB1}%
&&\hspace{-0.5cm}\delta_{K^{\prime}}^{\downarrow \uparrow} = - 2 m_{+} +\vert \lambda_{so} - m_{-} + V \vert + \vert \lambda_{so} - m_{-} - V \vert .
\label{eqB2}%
\end{eqnarray} 
For $m_b > \lambda_{so}$ and assuming $m_t < \lambda_{so}$, we find that
\begin{eqnarray}
\delta_{K}^{\downarrow \uparrow} &=& 2 ( \lambda_{so} - m_{t} ), \quad 0 < V < \lambda_{so} + m_{-},\\
\label{eqB3}%
\delta_{K}^{\downarrow \uparrow} &=& 2 ( V - m_{+} ),\quad \lambda_{so} + m_{-} < V.
\label{eqB4}%
\end{eqnarray}
In both cases $\delta_{K}^{\downarrow \uparrow} > 0$ because $m_{+} < \lambda_{so} + m_{-}$. For the gap $\delta_{K^{\prime}}^{\downarrow \uparrow}$ we find
\begin{eqnarray}
\hspace{-0.4cm}\delta_{K^{\prime}}^{\downarrow \uparrow} &=& 2 ( \lambda_{so} - m_{b} ) < 0, \quad 0 < V < \lambda_{so} - m_{-}, \\
\label{eqB5}%
\hspace{-4cm}\delta_{K^{\prime}}^{\downarrow \uparrow} &=& 2 ( V - m_{+} ), \quad \lambda_{so} - m_{-} < V < \lambda_{so} + m_{-}. 
\label{eqB6}%
\end{eqnarray}
Thus, we conclude that $\delta_{K^{\prime}}^{\downarrow \uparrow} < 0$ in the range $0 < V < m_{+}$. Note that $\lambda_{so} - m_{-} < m_{+} < \lambda_{so} + m_{-}$. Further, for $\lambda_{so} + m_{-} < V$, we find $\delta_{K}^{\downarrow \uparrow} = \delta_{K^{\prime}}^{\downarrow \uparrow} = 2 ( V - m_{+} ) > 0$.

\section{Velocity operator's matrix elements}

When $\lambda_R = 0$, the $x$ and $y$ components of the velocity operator $v_\nu = \partial H / \hbar \partial k_\nu$ ($\nu = x, y$) read
\begin{equation}
v_x = \tau v_F \sigma_x , \qquad
v_y = v_F \sigma_y .
\label{eqA2}%
\end{equation}
For the evaluation of the velocity matrix elements in Eq.~(\ref{eq22}) we introduce the notation
\begin{equation}
\langle u_{n k} \vert v_\nu \vert u_{n^{\prime} k} \rangle = v_{\nu, n n^{\prime}} ( k )  .
\label{eqA3}%
\end{equation}
The calculations are done for a specific valley and 
spin state. Using Eq.~(\ref{eq6}) they are readily evaluated and read
\begin{eqnarray}
v_{x, + -} ( k ) = - 
(\hbar v_F^2/\epsilon_k \varepsilon) \left(  i \varepsilon \tau k_y +M k_x  \right)  ,\\
\label{eqA4}%
v_{y, + -} ( k ) = 
(\hbar v_F^2/\epsilon_k \varepsilon) \left( i \varepsilon \tau k_x - M k_y \right)  .
\label{eqA5}%
\end{eqnarray}
\noindent Note that $v_{\nu, -+} ( k ) = v_{\nu, +-}^{\ast} ( k )$ due to the hermiticity of the velocity operator.

For $\lambda_R \neq 0$ the velocity operator reads
\begin{equation}
v_x = \tau v_F \sigma_x \mathbf{1}_s, \qquad 
v_y = v_F \sigma_y \mathbf{1}_s .
\label{eqA7}%
\end{equation}
In the following we suppress the spin chirality index. Using Eq.~(\ref{eq14}), the velocity matrix elements for the $K$ valley are readily evaluated and read
\begin{align}
\hspace{-0.7cm}
&v_{x, n n^{\prime}}^{K} ( k )   = P  
\Big[\Big( \frac{\epsilon_k \text{e}^{i \varphi_{k} } }{E_{ n k}-\gamma_1^{\uparrow}} + \frac{\epsilon_k \text{e}^{- i \varphi_{k } } }{E_{n^{\prime} k} -\gamma_1^{\uparrow}} \Big) \eta_{n k}^{\ast} \eta_{n^{\prime} k}   \nonumber
\\*
&\hspace{1.75cm} 
 + \frac{E_{n k} - \gamma_2^{\downarrow} }{\epsilon_k e^{- i \varphi_k}} + \frac{ E_{n^{\prime} k} - \gamma_2^{\downarrow} }{\epsilon_k e^{i \varphi_k}}
\Big] ,
\label{eqA8}%
\end{align}
where $P = N_{n k}^{\ast} N_{n^{\prime} k} v_F$ and 
\begin{align}
\hspace{-0.7cm}
&v_{y, n n^{\prime}}^{K} ( k )   = i P  
\Big[\Big( - \frac{\epsilon_k \text{e}^{i \varphi_{k} } }{E_{ n k}-\gamma_1^{\uparrow}} + \frac{\epsilon_k \text{e}^{- i \varphi_{k } } }{E_{n^{\prime} k} -\gamma_1^{\uparrow}} \Big) \eta_{n k}^{\ast} \eta_{n^{\prime} k}   \nonumber
\\*
&\hspace{1.75cm} 
 - \frac{E_{n k} - \gamma_2^{\downarrow} }{\epsilon_k e^{- i \varphi_k}} + \frac{ E_{n^{\prime} k} - \gamma_2^{\downarrow} }{\epsilon_k e^{i \varphi_k}}
\Big] .
\label{eqA9}%
\end{align}
Using Eq.~(\ref{eq15}), the velocity matrix elements for the $K^{\prime}$ valley are similarly evaluated and read
\begin{align}
\hspace{-0.7cm}
&v_{x, n n^{\prime}}^{K^{\prime}} ( k )   = \tilde{P}  
\Big[\Big( \frac{\epsilon_k \text{e}^{ - i \varphi_{k} } }{E_{ n^{\prime} k}-\gamma_2^{\uparrow}} + \frac{\epsilon_k \text{e}^{ i \varphi_{k } } }{E_{n k} -\gamma_2^{\uparrow}} \Big) \tilde{\eta}_{n k}^{\ast} \tilde{\eta}_{n^{\prime} k}   \nonumber
\\*
&\hspace{1.75cm} 
 + \frac{\epsilon_k e^{-i \varphi_k} }{ E_{n k} - \gamma_1^{\downarrow}} + \frac{ \epsilon_k e^{i \varphi_k} }{E_{n^{\prime} k} - \gamma_1^{\downarrow} }
\Big] ,
\label{eqA10}%
\end{align}
$\tilde{P} = \tilde{N}_{n k}^{\ast} \tilde{N}_{n^{\prime} k} v_F$, and 
\begin{align}
\hspace{-0.7cm}
&v_{y, n n^{\prime}}^{K^{\prime}} ( k )   = i \tilde{P}  
\Big[\Big( \frac{\epsilon_k \text{e}^{ - i \varphi_{k} } }{E_{ n^{\prime} k}-\gamma_2^{\uparrow}} - \frac{\epsilon_k \text{e}^{ i \varphi_{k } } }{E_{n k} -\gamma_2^{\uparrow}} \Big) \tilde{\eta}_{n k}^{\ast} \tilde{\eta}_{n^{\prime} k}   \nonumber
\\*
&\hspace{1.75cm} 
 + \frac{\epsilon_k e^{-i \varphi_k} }{ E_{n k} - \gamma_1^{\downarrow}} - \frac{ \epsilon_k e^{i \varphi_k} }{E_{n^{\prime} k} - \gamma_1^{\downarrow} }
\Big] .
\label{eqA11}%
\end{align}
The above matrix elements are used in the calculation of the Berry curvatures.

\end{appendix}


\begin{thebibliography}{9}
\bibitem{comment2} We use the definition of the spin Chern number as derived from a $2\times2$ Chern number matrix $\mathcal{C}^{\alpha,\beta}$ with $\alpha,\beta = \uparrow,\downarrow$, proposed by Sheng and collaborators \cite{sheng06}. Particularly, the spin Chern number was defined by integrating the Berry curvature generated by imposing twisted boundary conditions on a finite sample. In the case of zero Rashba SOC, only the diagonal matrix elements are nonzero and given by $\mathcal{C^{\uparrow,\uparrow}} = \nu_{\uparrow} = 1$ and $\mathcal{C^{\downarrow,\downarrow}} = \nu_{\downarrow} = -1$. Consequently, the spin Chern number was shown to be quantized to $\mathcal{C}_s = \sum_{\alpha,\beta} \alpha \mathcal{C}^{\alpha,\beta} = \nu_\uparrow - \nu_{\downarrow} = 2$. The spin-Hall conductivity is then given as $\sigma_{xy}^{s} = \mathcal{C}_s = 2$ in units of $e/4 \pi$. The spin Chern number has also been defined as the half difference between the Chern numbers for the spin up and spin down sectors of the occupied space, $\mathcal{C}_s = \frac{1}{2} \left ( \mathcal{C}_\uparrow - \mathcal{C}_\downarrow \right )$, which was derived for an infinite sample in the thermodymic limit without using any boundary conditions \cite{prodan09} and is quantized to $\mathcal{C}_s = 1$. Both definitions are equally valid descriptions of the spin Chern number and they both imply that $\mathcal{C}_s$ is a robust topological invariant. However, in the first definition the spin-Hall conductivity is given as $\sigma_{xy}^s = (e/4 \pi)\mathcal{C}_s$, whereas in the second definition $\sigma_{xy}^s = (e/2\pi) \mathcal{C}_s$. Both definitions yield $\sigma_{xy}^s = e/2\pi$.
\end{thebibliography}
\end{document}